\documentclass[journal,onecolumn,draftcls,12pt]{IEEEtran}
\usepackage[T1]{fontenc}
\pdfoutput=1

\makeatletter
\def\endthebibliography{%
	\def\@noitemerr{\@latex@warning{Empty `thebibliography' environment}}%
	\endlist
}
\makeatother

\IEEEoverridecommandlockouts
\usepackage{cite}
\usepackage{amsmath,amssymb,amsfonts}
\usepackage{algorithm}
\usepackage{algorithmic}
\usepackage{etoolbox}  
\usepackage{bbm}
\usepackage{subfig}
\usepackage{amsmath}
\usepackage{dsfont}
\usepackage{tabularx}
\usepackage{amssymb}
\usepackage{threeparttable}
\usepackage{booktabs}
\usepackage{amsthm}
\usepackage{mathtools}
\usepackage{bm}
\usepackage{url}
\makeatletter
\patchcmd{\algorithmic}{\addtolength{\ALC@tlm}{\leftmargin} }{\addtolength{\ALC@tlm}{\leftmargin}}{}{}
\makeatother

\makeatletter
\newcommand\fs@betterruled{%
	\def\@fs@cfont{\bfseries}\let\@fs@capt\floatc@ruled
	\def\@fs@pre{\vspace*{5pt}\hrule height.8pt depth0pt \kern2pt}%
	\def\@fs@post{\kern2pt\hrule\relax}%
	\def\@fs@mid{\kern2pt\hrule\kern2pt}%
	\let\@fs@iftopcapt\iftrue}
\floatstyle{betterruled}
\restylefloat{algorithm}
\makeatother

\usepackage{tikz}
\usepackage{pgfplots}
\usepackage{tikzscale}
\usepackage{tikz-qtree}
\pgfplotsset{compat=newest}
\pgfplotsset{plot coordinates/math parser=false}
\pgfplotsset{every axis/.append style={
                    label style={font=\scriptsize},
                    tick label style={font=\scriptsize},
                    legend style={font=\scriptsize}
                    }}
\usetikzlibrary{plotmarks,shapes,patterns,decorations.pathreplacing,backgrounds,calc,arrows,arrows.meta,spy,matrix,shadows,trees,positioning,fit}
\usepgfplotslibrary{patchplots,groupplots,units}

\tikzstyle{startstop} = [rectangle, rounded corners, minimum width=2cm, minimum height=0.5cm,text centered, draw=black]
\tikzstyle{io} = [trapezium, trapezium left angle=70, trapezium right angle=110, minimum width=3cm, minimum height=1cm, text centered, draw=black]
\tikzstyle{process} = [rectangle, minimum width=2cm, minimum height=0.5cm, text centered, draw=black, alignb=center]
\tikzstyle{decision} = [ellipse, minimum width=2cm, minimum height=1cm, text centered, draw=black]
\tikzstyle{arrow} = [thick,<->,>=stealth]
\tikzstyle{line} = [thick,>=stealth]
\tikzstyle{darrow} = [thick,<->,>=stealth,dashed]
\tikzstyle{sarrow} = [thick,->,>=stealth]
\tikzstyle{larrow} = [line width=0.1mm,dashdotted,->,>=stealth]

\pgfkeys{/pgf/number format/.cd,1000 sep={\,}} 

\makeatletter
\def\grd@save@target#1{%
  \def\grd@target{#1}}
\def\grd@save@start#1{%
  \def\grd@start{#1}}
\tikzset{
  grid with coordinates/.style={
    to path={%
      \pgfextra{%
        \edef\grd@@target{(\tikztotarget)}%
        \tikz@scan@one@point\grd@save@target\grd@@target\relax
        \edef\grd@@start{(\tikztostart)}%
        \tikz@scan@one@point\grd@save@start\grd@@start\relax
        \draw[minor help lines] (\tikztostart) grid (\tikztotarget);
        \draw[major help lines] (\tikztostart) grid (\tikztotarget);
        \grd@start
        \pgfmathsetmacro{\grd@xa}{\the\pgf@x/1cm}
        \pgfmathsetmacro{\grd@ya}{\the\pgf@y/1cm}
        \grd@target
        \pgfmathsetmacro{\grd@xb}{\the\pgf@x/1cm}
        \pgfmathsetmacro{\grd@yb}{\the\pgf@y/1cm}
        \pgfmathsetmacro{\grd@xc}{\grd@xa + \pgfkeysvalueof{/tikz/grid with coordinates/major step x}}
        \pgfmathsetmacro{\grd@yc}{\grd@ya + \pgfkeysvalueof{/tikz/grid with coordinates/major step y}}
        \foreach \x in {\grd@xa,\grd@xc,...,\grd@xb}
        \node[anchor=north] at (\x,\grd@ya) {\pgfmathprintnumber{\x}};
        \foreach \y in {\grd@ya,\grd@yc,...,\grd@yb}
        \node[anchor=east] at (\grd@xa,\y) {\pgfmathprintnumber{\y}};
      }
    }
  },
  minor help lines/.style={
    help lines,
    gray,
    line cap =round,
    xstep=\pgfkeysvalueof{/tikz/grid with coordinates/minor step x},
    ystep=\pgfkeysvalueof{/tikz/grid with coordinates/minor step y}
  },
  major help lines/.style={
    help lines,
    line cap =round,
    line width=\pgfkeysvalueof{/tikz/grid with coordinates/major line width},
    xstep=\pgfkeysvalueof{/tikz/grid with coordinates/major step x},
    ystep=\pgfkeysvalueof{/tikz/grid with coordinates/major step y}
  },
  grid with coordinates/.cd,
  minor step x/.initial=.5,
  minor step y/.initial=.2,
  major step x/.initial=1,
  major step y/.initial=1,
  major line width/.initial=1pt,
}
\makeatother

\newlength\fheight
\newlength\fwidth

\makeatletter
\def\endthebibliography{%
	\def\@noitemerr{\@latex@warning{Empty `thebibliography' environment}}%
	\endlist
}
\makeatother

\usepackage{adjustbox}

\usepackage{gincltex}
\usepgfplotslibrary{fillbetween}
\usetikzlibrary{plotmarks}
\usetikzlibrary{patterns}
\usepackage{graphicx}
\usepackage{textcomp}
\usepackage{xcolor}
\usepackage{authblk}
\def\BibTeX{{\rm B\kern-.05em{\sc i\kern-.025em b}\kern-.08em
		T\kern-.1667em\lower.7ex\hbox{E}\kern-.125emX}}

\usepgfplotslibrary{external}
\pgfplotsset{compat=1.15}

\usepackage{glossaries}

\newacronym{ai}{AI}{Artificial Intelligence}
\newacronym{urllc}{URLLC}{Ultra-Reliable Low-Latency Communications}
\newacronym{qos}{QoS}{Quality of Service}
\newacronym{pdf}{PDF}{Probability Density Function}
\newacronym{pmf}{PMF}{Probability Mass Function}
\newacronym{cdf}{CDF}{Cumulative Density Function}
\newacronym{iot}{IoT}{Internet of Things}
\newacronym{aoi}{AoI}{Age of Information}
\newacronym{paoi}{PAoI}{Peak AoI}
\newacronym{pec}{PEC}{Packet Erasure Channel}
\newacronym{foi}{FoI}{Frequency of Information}
\newacronym{lifo}{LIFO}{Last In First Out}
\newacronym{fifo}{FIFO}{First In First Out}
\newacronym{mdp}{MDP}{Markov Decision Process}
\newacronym{pi}{PI}{Policy Iteration}
\newacronym{rl}{RL}{Reinforcement Learning}
\newacronym{mec}{MEC}{Mobile Edge Computing}
\newacronym{gps}{GPS}{Generalized Processor Sharing}
\newacronym{ar}{AR}{Augmented Reality}
\newacronym{jfi}{JFI}{Jain Fairness Index}
\newacronym{xr}{XR}{eXtended Reality}
\newacronym{nfv}{NFV}{Network Function Virtualization}
\newacronym{aop}{AoP}{Age of Processing}

\newcommand{\E}[1]{\mathbb{E}\left[ #1 \right]} 
\newcommand{\mc}[1]{\mathcal{#1}}   
\newcommand{\mb}[1]{\mathbf{#1}}    

\def \fwidth{0.9\columnwidth}
\def \fheight{0.45\columnwidth}
\def \sfwidth{0.48\linewidth}
\def \sfheight {0.32\linewidth}
\def \tfwidth{0.32\linewidth}

\definecolor{color0}{HTML}{FFD700}
\definecolor{color1}{HTML}{FFB14E}
\definecolor{color2}{HTML}{FA8775}
\definecolor{color3}{HTML}{EA5F94}
\definecolor{color4}{HTML}{CD34B5}
\definecolor{color5}{HTML}{9D02D7}
\definecolor{color6}{HTML}{0000FF}

\begin{document}

\title{Age of Information Analysis in Shared Edge Computing Servers}

\author{Federico Chiariotti,~\IEEEmembership{Member,~IEEE}\thanks{Federico Chiariotti (chiariot@dei.unipd.it) is with the Department of Information Engineering, University of Padova, Italy. This work was supported by the European Union under the Italian National Recovery and Resilience Plan of NextGenerationEU, as part of the Young Researchers grant ``REDIAL'' (SoE0000009).}
} 

\maketitle

\begin{abstract}
\gls{mec} is expected to play a significant role in the development of 6G networks, as new applications such as cooperative driving and \gls{xr} require both communication and computational resources from the network edge. However, the limited capabilities of edge servers may be strained to perform complex computational tasks within strict latency bounds for multiple clients. In these contexts, both maintaining a low average \gls{aoi} and guaranteeing a low \gls{paoi} even in the worst case may have significant user experience and safety implications. In this work, we investigate a theoretical model of a \gls{mec} server, deriving the expected \gls{aoi} and the \gls{paoi} and latency distributions under the \gls{fifo} and \gls{gps} resource allocation policies. We consider both synchronized and unsynchronized systems, and draw insights on the robust design of resource allocation policies from the analytical results.
\end{abstract}

\IEEEpeerreviewmaketitle
\glsresetall

\section{Introduction}\label{sec:intro}

One of the promises of the upcoming sixth generation of mobile networks is the union of communication, computation, and sensing as a single service provided by the network~\cite{qi2022integrating}: in particular, the rapid developments in \gls{ai} generated a significant demand for low-latency computing power~\cite{letaief2021edge}, which must be placed at the network edge to avoid long propagation times~\cite{tr28814}.

Cooperative driving is one of the major examples of this trend~\cite{sun2018cooperative}:
automated vehicles require huge neural networks to integrate the data from disparate sensors into a coherent picture, and the emissions from on-board computers may become a significant carbon emitter over the next decade~\cite{sudhakar2022data}. The efficiency gains from deploying \gls{mec}, and the possibility of coordinating vehicles in an area~\cite{balkus2022survey}, has led to a significant amount of research on computational offloading~\cite{liu2021vehicular} in this context. However, the stringent safety requirements of autonomous vehicles pose a significant challenge: the sensory data from multiple vehicles must be processed and the results of the computation must be distributed within very strict times, often measured in milliseconds. The computational capabilities of edge nodes, or even more mobile platforms such as drones~\cite{traspadini2022uav}, may be hard-pressed to meet these requirements, and system optimization is non-trivial.

Another significant use case for computational offloading is represented by \gls{xr}~\cite{tr38838}: headsets must be battery-powered and light not to encumber the user, but the computational requirements for rendering high-definition virtual scenes, or even overlaying objects on the user's field of view, are significant~\cite{morin2022toward}. The unpredictable nature of \gls{xr} traffic poses further challenges~\cite{itu-t-f.743.10}, and the IEEE set a maximum motion-to-photon latency of 20~ms~\cite{ieee2021vrstandard} in a recent standard, in order to avoid the insurgence of \emph{cybersickness}, a condition similar to motion sickness caused by a mismatch between visual and proprioceptive sensory inputs.

Computational resource allocation then becomes a significant challenge, which has attracted a significant amount of interest from the research and industrial communities~\cite{zhao2019computation}. \gls{gps} is a classical policy~\cite{demers1989analysis,parekh1994generalized} that is known to maximize fairness between clients, and is often used in \gls{nfv} and task offloading~\cite{tang2020deep,zhang2021risk}: it runs all outstanding tasks in parallel by equally splitting the computational resources among them. The opposite policy, \gls{fifo} queuing, allocates all available resources to the client that started service first, providing a faster service once the clients get to the head of the queue but requiring a waiting time if the server is busy. These ubiquitous policies provide a benchmark for any more complex resource allocation mechanisms, and often work extremely well in practical scenarios in terms of latency. However, latency may not capture the full picture in the most relevant use cases~\cite{popovski2022perspective}: since the processed data are used for control in both the \gls{xr} and vehicular use cases, the \gls{aoi}, i.e., the freshness of the information available to the \gls{xr} user or the self-driving vehicle when they make a decision, is a more compact and meaningful metric~\cite{chiariotti2021peak}. In particular, \gls{paoi} represents an upper bound to the age, and can be used in reliability contexts. These metrics have been extensively used in \gls{xr}~\cite{chaccour2020ruin,franco2019reliability} and vehicular~\cite{abdel2019optimized} use cases.

\begin{figure*}[t!]
    \centering
\input{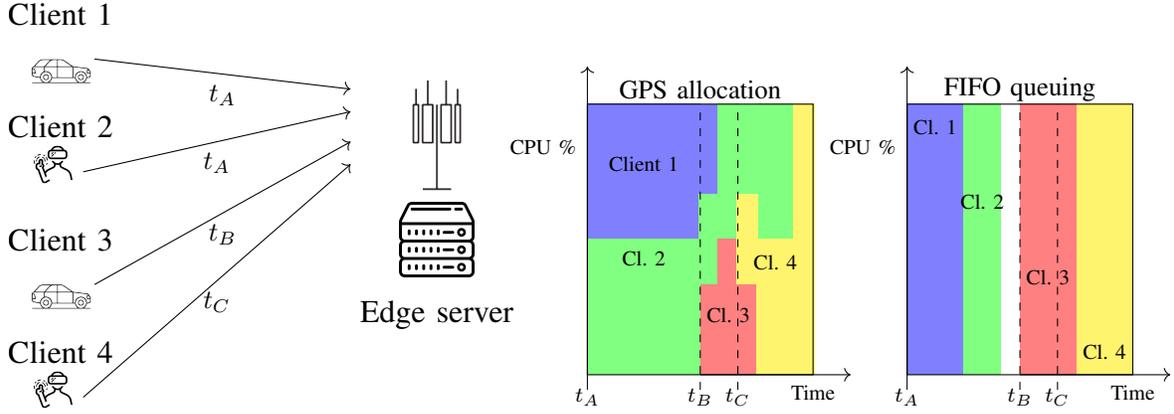}
    \caption{Depiction of the scenario and of a possible outcome of the two resource sharing policies.}
    \label{fig:scenario}
\end{figure*}

In this work, we model a scenario such as the one in Fig.~\ref{fig:scenario}, in which several clients must share the limited computational resources of a \gls{mec} server, and older frames are preempted by new requests from the same client. A possible example of the outcomes of the two opposite policies is also shown in the figure: as described above, \gls{gps} divides the available capacity among all active clients, serving them simultaneously at a fraction of the full capacity, while \gls{fifo} divides clients in time, giving each one the full capacity of the server, one after the other. We consider the periodic generation of equally complex computational tasks and analyze the distribution of the \gls{paoi}, drawing insights on the design of such a system. In particular, our contributions are:
\begin{itemize}
 \item We analyze the synchronized case, in which users generate frames at the same time, and provide closed-form expressions for the expected \gls{aoi} and for the \glspl{cdf} of the computing latency and \gls{paoi} for a given client;
 \item We extend the analysis to the general case in which clients generate frames in batches, under the \gls{fifo} and \gls{gps} policies, deriving the distributions of the latency and \gls{paoi}, as well as the expected \gls{aoi};
 \item We verify the analysis by Monte Carlo simulation\footnote{All the code necessary to reproduce the theoretical and Monte Carlo results is publicly available in the following repository: \url{https://github.com/signetlabdei/aoi_edge_computing}} and provide design guidelines for \gls{mec} systems, analyzing the robustness of the policies to the drift between clients' clocks.
\end{itemize}

To the best of our knowledge, this work is the first to theoretically model this scenario, which is extremely relevant for future 6G use cases, and derive the metrics analytically. The rest of the paper is organized as follows: first, Sec.~\ref{sec:related} presents an overview of the state of the art on the subject. The system model is then described in Sec.~\ref{sec:sys}, and the analyses for the synchronized and general case are presented in Sec.~\ref{sec:sync} and Sec.~\ref{sec:batch}, respectively. Sec.~\ref{sec:results} then presents the simulations verifying the theoretical calculations, along with the possible design insights deriving from the results, and finally, Sec.~\ref{sec:conc} concludes the paper.

\section{Related Work}\label{sec:related}

As we mentioned in the Introduction, placing computational power as close as possible to users is a major trend in mobile networks~\cite{letaief2021edge}, enabling almost real-time processing of complex data such as \gls{xr} frames or vehicular sensory information. The concept of network slicing, i.e., dynamically allocating communication and computational resources based on the \gls{qos} requirements of users, is another pillar of the evolution of mobile networks into 5G~\cite{zhang2019overview} and beyond~\cite{wu2022ai}. Thirdly, the importance of \gls{aoi} in remote monitoring and control systems has been recognized since the metric's inception~\cite{kaul2012real}, but most of the research community's focus was on communications. \gls{aoi}, and related metrics such as \gls{paoi}, have only been applied more recently~\cite{kuang2019age} to edge computing cases, but age-aware slicing policies are crucial for the novel 6G applications mentioned above.

The joint optimization of wireless transmission and computing aspects is a complex task that is often modeled as a \gls{mdp}~\cite{jayanth2022age} in the relevant literature: as the arrival process to the edge server depends on the access mechanism and wireless channel properties, computing the tandem queue delay and age can require multiple operations, and the complexity of optimization scales accordingly. Considering the worst-case \gls{paoi}, in terms of either the violation probability for a specific requirement or a high percentile of the distribution, also increases the complexity of the calculation, but the closed-form \gls{paoi} distribution has been derived for some simple systems~\cite{chiariotti2021peak}. The scheduling of updates in this context can lead to better performance~\cite{zhu2022online}, but requires the sensor to be able to receive and process feedback, with a higher energy consumption, which may be offset by harvesting energy from the environment~\cite{liu2021optimizing}.

Existing works also consider the effect of different offloading policies between the Edge and Cloud on the \gls{aoi}~\cite{qin2022timeliness}: the massive processing capabilities of the Cloud reduce the load, but are partially offset by the higher transmission delay~\cite{zou2021timely}. The choice between offloaded and local processing is also difficult~\cite{he2022age}, as the features of the Edge server~\cite{traspadini2022uav} and the traffic load on it~\cite{li2022analysis} may significantly affect the resulting \gls{aoi}. This case is also modeled as an \gls{mdp} to determine optimal scheduling, as stateful offloading decisions manage to keep a lower load and reduce the age~\cite{jiang2023age}. The same models may be used to consider sensing~\cite{chen2023joint} or compression~\cite{hu2022age} before transmission under different policies, which may also affect access to the shared wireless channel~\cite{zhou2021age}.

This type of analysis has been applied to several realistic scenarios. In this case, costs are often defined on the system level, i.e., in terms of operating costs or provider-level \gls{qos} metrics. The optimization of Smart City processing is considered in~\cite{modina2022joint}, moving computational functions across cells and to the Cloud when necessary, while the radio access using OpenRAN in a vehicular scenario is considered in~\cite{ndikumana2022age}. Industrial system may also benefit from age-aware policies~\cite{xie2023scheduling}, as the efficiency and energy consumption of control and computation offloading can be improved by considering \gls{aoi} explicitly~\cite{huang2023aoi}. Finally, the \gls{aoi} in mixed air-ground networks is considered in~\cite{chen2021information} using game theoretical concepts to model the interactions between individual users and computing-enabled flying base stations.

The research community's interest in freshness-based analysis in mixed communication and computational systems is growing, and some recent works have even started using \gls{aop}~\cite{li2021age} to refer to the age in computational offloading scenarios, to highlight that the system does not just relay and monitor information, but rather needs to actively perform computational transformations~\cite{liu2022microservice}. In this context, this work aims at a rigorous analysis of common processor sharing policies, both in terms of average \gls{aoi} and worst-case \gls{paoi}, so as to provide a baseline for future resource slicing algorithms based on more advanced stateful policies.

\section{System Model}\label{sec:sys}

Let us consider a scenario in which $N$ clients need to share a \gls{mec} server's computational capacity for tasks with strict timeliness constraints: possible applications include the rendering of users' different viewpoints in a shared \gls{ar} environment, or the processing of camera or LIDAR frames generated by autonomous vehicles approaching a crossing. In both cases, each client generates \emph{frames} with a fixed period $\tau$. The \gls{mec} server then processes those frames; we assume that all frames require a stochastic time to be fully processed, following an exponential distribution with rate $\mu$.

In the following, we will consider two different ways of allocating the server's computational capabilities:
\begin{itemize}
 \item Frames can be processed \emph{sequentially}, following a \gls{fifo} queuing model: the client with the oldest frame in service is served first, with the full capabilities of the server allocated exclusively to it;
 \item Frames can be served \emph{in parallel}, following the well-known \gls{gps} policy. In this case, if there are $X\leq N$ active clients being served, each of their frames will be processed with rate $\frac{\mu}{X}$.
\end{itemize}
In both cases, we consider a preemptive system, i.e., if frame $i$ from client $n$ is still being processed after a time $\tau$, the $i+1$-th frame from that client supersedes it, so the older frame is discarded and the newer version replaces it in service or in the queue. This assumption is widely used in the \gls{aoi} literature, and fits systems such as the ones in the examples, as the dynamic nature of the environment makes older data rapidly useless.

In the following, we consider two cases for our optimization:
\begin{itemize}
 \item In a \emph{synchronized} system, all clients generate frames at the same time: the $i$-th frame for client $n$ is generated at time $g_{i,n}=i\tau\,\forall n\in\{1,\ldots,N\}$;
 \item In a \emph{batch} system, there are $B$ batches of clients generating frames at the same time, with an offset $\nu_b$ from the previous batch. Each batch $b$ contains $N_b$ clients, with $\sum_{b=1}^B N_b=N$. The $i$-th frame from a client $n$ belonging to batch $b$, we then have $g_{i,n}=i\tau+\sum_{j=2}^b\nu_b$. Naturally, we also have $\sum_{b=2}^B\nu_b\leq\tau$.
\end{itemize}
The synchronized system is a special case of a batch system with $B=1$; the opposite extreme, with $B=N$, corresponds to a case in which all clients generate frames at different times. In the \gls{fifo} scheme, we consider frames arriving simultaneously to be queued randomly every time, as minimal offsets in the clocks of different clients may result in different orderings.

We then define the timeliness metrics we consider in evaluating the two systems: let us consider a single client $n$. The $i$-th frame from that client is generated at time $g_{i,n}$, and the instant in which it is served is denoted as $s_{i,n}$, with $s_{i,n}=+\infty$ if the frame is dropped. The latency $T_{i,n}$ for the $i$-th frame from client $n$ is then simply given by:
\begin{equation}
 T_{i,n}=s_{i,n}-g_{i,n}.
\end{equation}
We also define the set $\mc{D}_n(t)$, which includes all the frames from client $n$ that have been served by time $t$:
\begin{equation}
 \mc{D}_n(t)=\left\{i\in\mathbb{N}:s_{i,n}\leq t\right\}.
\end{equation}
We can then define the \gls{aoi} $\Delta(t)$ as the freshness of the latest received packet at time $t$, i.e., the time that has elapsed since its generation:
\begin{equation}
 \Delta(t)=t-\max_{i\in\mc{D}_n(t)}g_{i,n}.
\end{equation}
The average \gls{aoi} $\bar{\Delta}$ is defined as follows:
\begin{equation}
 \bar{\Delta}=\lim_{T\rightarrow\infty}\frac{1}{T}\int_0^T\Delta(t)dt.
\end{equation}
We can also define the \gls{paoi} $\Psi_{i,n}$ for each successfully delivered frame:
\begin{equation}
 \Psi_{i,n}=s_{i,n}-\max_{j\in\mc{D}_n(s_{i,n})}g_{j,n},\text{ if }\exists T\in\mathbb{R}:s_{i,n}\leq T.
\end{equation}
We can then use the latency and \gls{paoi} as metrics for the timeliness of the \gls{mec} offloading service, computing their complete distribution, as well as the expected \gls{aoi} $\bar{\Delta}$.

\section{Analysis: Synchronized Frame Generation}\label{sec:sync}

We first consider the synchronized case in which frames are generated simultaneously by each client and they all have the same period $\tau$. In this case, the \gls{mec} server is always fully booked right after a frame is generated. On the other hand, the \gls{pmf} of the number of clients right before a new frame is generated, $X_{\tau-\varepsilon}$, where $\varepsilon>0$ is an infinitesimal value, is given by:
\begin{equation}
 p_{X_{\tau-\varepsilon}}(k)=\begin{cases}
                            \frac{\gamma_N(\mu\tau)}{(N-1)!} &\text{if }k=0;\\
                            \frac{(\mu\tau)^{N-k}e^{-\mu\tau}}{(N-k)!} &\text{if }k\in\{1,\ldots,N\},
                            \end{cases}
\end{equation}
where $\gamma_k(x)$ is the lower incomplete gamma function, which is defined as follows:
\begin{equation}
 \gamma_k(x)=\int_0^x e^{-t}t^{k-1}dt.
\end{equation}

In this special case, there is no prioritization or ordering, as clients are queued randomly, and considering a \gls{gps} system or a \gls{fifo} queue (which corresponds to a $D^N/M/1$ in the standard Kendall notation) is then exactly the same thanks to the properties of Poisson processes. As all clients are identical, the success probability $\sigma$ of a frame from a given client is given by:
\begin{equation}
 \sigma=\sum_{k=0}^{N-1} \frac{N-k}{N} p_{X_{\tau-\varepsilon}}(k).
\end{equation}
Naturally, if $X_{\tau-\varepsilon}=N$, no frames have been served by the \gls{mec} and no client successfully managed to deliver a frame. As the queue is preemptive, i.e., frames that are not successful before the next frames are generated are substituted by the newer ones, this repeats identically for each frame.

In order to compute the delay experienced by a specific (successful) frame, we can simply consider the fact that, in order for a specific client to have been served by time $t$, we need to satisfy two conditions: firstly, at least $k\geq1$ frames must have been served by time $t$, and secondly, that specific client must be among them. As clients are all identical, the probability of being successful if $k$ frames have been served is simply $\frac{k}{N}$. We can then apply the law of total probability, knowing that served frames follow a Poisson process:
\begin{equation}
 P_T(t)=\frac{\gamma_N(\mu t)}{\sigma(N-1)!}+\sum_{k=1}^{N-1}\frac{(\mu t)^k e^{-\mu t}}{N(k-1)!}.\label{eq:sync_PT}
\end{equation}
Naturally, we have $P_T(\tau)=1$, as this only consider frames successfully delivered by time $\tau$. The \gls{cdf} of the \gls{paoi} $\Psi$ is then given by:
\begin{equation}
 P_{\Psi}(\psi)=\!\left(\!1-(1-\sigma)^{\left\lfloor\frac{\psi-\tau}{\tau}\right\rfloor}(1-\sigma P_T(\text{mod}(\psi,\tau)))\!\right)\!u(\psi-\tau),
\end{equation}
where $\text{mod}(m,n)$ is the integer modulo function and $u(x)$ is the stepwise function, equal to 1 if $x\geq 0$ and 0 otherwise.

We can also compute the average \gls{aoi} following the well-known geometric method~\cite{yates2021age}: if the interval between the generation of two successful frames is $Y$, and the latency of the latter is $T$, the average \gls{aoi} $\bar{\Delta}$ can be computed as:
\begin{equation}
 \bar{\Delta}=\frac{1}{\E{Y}}\left(\frac{\E{(T+Y)^2}}{2}-\frac{\E{T^2}}{2}\right)
 =\frac{\E{Y^2}}{2\E{Y}}+\frac{\E{TY}}{\E{Y}}.\label{eq:trapezoids}
\end{equation}
As the latency of a frame is independent from what happened before its generation, we know that $\E{TY}=\E{T}\E{Y}$, so we get:
\begin{equation}
 \bar{\Delta}=\frac{\E{Y^2}}{2\E{Y}}+\E{T}.
\end{equation}
We know that $Y$ follows a geometric distribution with probability $\sigma$, so that $\E{Y}=\tau \sigma^{-1}$ and $\E{Y^2}=\frac{\tau^2(2-\sigma)}{\sigma^2}$. The expected value of the latency can be computed by integrating $1-P_T(t)$, as given by~\eqref{eq:sync_PT}:
\begin{equation}
 \begin{aligned}
  \E{T}=&\int_0^{\tau}1-\frac{\gamma_N(\mu t)}{\sigma(N-1)!}-\sum_{k=1}^{N-1}\frac{(\mu t)^k e^{-\mu t}}{N(k-1)!}dt=-\frac{(1-\sigma)\tau}{\sigma}+\sum_{k=0}^{N-1}\frac{(N-k)\gamma_{k+1}(\mu\tau)}{\mu \sigma N k!}.\label{eq:T_sync}
 \end{aligned}
\end{equation}
We can then get the average \gls{aoi}:
\begin{equation}
 \begin{aligned}
  \bar{\Delta}=&\frac{(2-\sigma)\tau}{2\sigma}+\E{T} =\frac{\tau}{2}+\sum_{k=0}^{N-1}\frac{(N-k)\gamma_{k+1}(\mu\tau)}{\mu \sigma N k!}.
 \end{aligned}
\end{equation}
The computational complexity to compute the latency and \gls{paoi} \glspl{cdf}, or the average \gls{aoi}, is always $O(N)$.

\section{Analysis: Batch Frame Generation}\label{sec:batch}
We can now consider the more complex case in which clients are grouped in $B$ batches. In order to simplify the notation, we group the time offset for each batch $b$ in vector $\bm{\nu}$, where $\nu_b$ is the time that occurs between the generation of batches $b-1$ and $b$, and $\nu_1=\tau-\sum_{b=2}^B\nu_b$.

\subsection{GPS Operation}

We first analyze the \gls{gps} system. We can consider the instants immediately before the generation of each batch as the steps of a Markov process, where the state $\mb{s}$ is given by:
\begin{itemize}
 \item The index $s_0$ of the next batch;
 \item The number $s_b\in\{0,\ldots,N_b\}$ of frames from each batch $b$ that are still in processing immediately before the next batch is generated, i.e., at time $\nu_{s_0}-\varepsilon$.
\end{itemize}
The state vector is then $\mb{s}=\left(s_0,s_1,\ldots,s_B\right)$, and the state space of the Markov chain representing the system is $\{1,\ldots B\}\times\prod_{b=1}^B\{0,\ldots,N_b\}$. In the most extreme case, $B=N$ and all clients arrive at different times: the size of the Markov chain in that case is $N2^N$. For the ease of the reader, we also define subset $\mc{S}_b$ as follows:
\begin{equation}
 \mc{S}_b=\left\{\mb{s}\in\mc{S}:s_0=b\right\}.\label{eq:sb}
\end{equation}

A transition from state $\bm{s}$ to state $\bm{s}'$ is possible if two conditions are met: firstly, $s_0'=s_0+1$, as the batch indexes are traversed sequentially.\footnote{With a slight abuse of notation, we allow batch indices larger than $B$ in equations, considering $b>B$ to be equivalent to writing $\text{mod}(b,B)$ in full. This should be apparent to the reader, and allows for more compact and readable equations.} Secondly, aside from batch $s_0$, there are no new arrivals, so $s_b'\leq s_b,\,\forall b\neq s_0$. We denote the set of states that are reachable from $\bm{s}$ as $\mc{R}(\bm{s})$:
\begin{equation}
 \mc{R}(\bm{s})=\left\{\bm{s}'\in\mc{S}_{s_0+1}: s_b'\leq s_b,\,\forall b\notin\{0,s_0\}\right\}.
\end{equation}
Additionally, we define the number of active frames right after state $\bm{s}$, $X(\bm{s})$, as follows:
\begin{equation}
 X(\bm{s})=N_{s_0}+\sum_{b=1,b\neq s_0}^B s_b.
\end{equation}
The value includes all the frames still in processing from other batches, summed to the frames in the new batch that just arrived.
We then give the \gls{pmf} of the number $C$ of completed frames in interval $t$, knowing the initial state and that no new frames are generated before time $t$:
\begin{equation}
 p_{C|S,T}(c|\bm{s},t)=\begin{cases}
 \frac{(\mu t)^c e^{-\mu t}}{c!} &\text{if }c<X(\bm{s});\\
                            \frac{\gamma_{X(\bm{s})}(\mu t)}{(X(\bm{s})-1)!} &\text{if }c=X(\bm{s}).
                            \end{cases}\label{eq:events_state}
\end{equation}
If we consider that $c$ frames were successfully processed over the interval, the transition probability to the next state follows a multivariate hypergeometric distribution. The transition probability is then given by:
\begin{equation}
 P(\bm{s},\bm{s}')=\begin{cases}
  p_{C|S,T}\left(X(\bm{s})-\sum_{b=1}^B s_b'\Big|\bm{s},\nu_{s_0+1}\right)\frac{\binom{N_{s_0}}{N_{s_0}-s'_{s_0}}\prod_{b=1,b\neq s_0}^B \binom{s_b}{s_b-s'_b}}{\binom{X(\bm{s})}{X(\bm{s})-\sum_{b=1}^B s_b'}}, &\text{if }\bm{s}'\in\mc{R}(s);\\
  0, &\text{otherwise.}
 \end{cases}
\end{equation}

We can then define a transition matrix $\mb{P}$, whose entries correspond to the transition probabilities. As all frames are equally likely to be completed, the transition probability is uniform among states that correspond to the same number of events between the two batches. The transition probability matrix over $m$ steps is simply $\mb{P}^m$, following the Markov property. The success probability $\sigma(\bm{s})$ for a client generated when in state $s$ is then given by:
\begin{equation}
 \sigma(\bm{s})=\sum_{\bm{s}'\in\mc{S}_{s_0}}P^B(\bm{s},\bm{s}')\frac{N_{s_0}-s'_{s_0}}{N_{s_0}}.
\end{equation}
Using the transition probability matrix, we can then easily compute the steady-state probability vector $\bm{\pi}$ by using the eigenvalue method. The success probability for a client arriving in batch $b$ is then given by:
\begin{equation}
 \sigma_b=\sum_{\bm{s}\in\mc{S}_b}\frac{\pi(\bm{s})\sigma(\bm{s})}{\sum_{\bm{s}'\in\mc{S}_b}\pi(\bm{s}')}.\label{eq:tot_success}
\end{equation}
In order to derive the latency \gls{cdf}, we first define the probability $P_{D|X}(t|x)$ of a specific client being served in a time interval $t$, if there are $x$ frames in the system at time $0$ and no new frames arrive. This is simply equal to the latency \gls{cdf} in the synchronized system, as defined in~\eqref{eq:sync_PT}.

We also define the batch index $\beta_b(t)$, which corresponds to the last batch to be generated before a time $t$ has passed since batch $b$:
\begin{equation}
 \beta_b(t)=\inf\left\{i\in\{b,\ldots,B+b-1\}:\sum_{k=b}^i \nu_{k+1}\geq t\right\}.\label{eq:batch_idx}
\end{equation}
We can also define the batch time $\tau(b,b')$ as the time between batch $b$ and the first instance of batch $b'$:
\begin{equation}
 \tau(b,b')=\begin{cases}
        \sum_{i=b+1}^{b'}\nu_i,                       &\text{if }b'>b;\\
        \sum_{i=b+1}^{B+b'}\nu_i,     &\text{if }b'<b;\\
        0,                                          &\text{if }b=b'.
        \end{cases}\label{eq:batch_time}
\end{equation}

We can then combine~\eqref{eq:sync_PT} and~\eqref{eq:batch_idx} to get the \gls{cdf} of the latency, considering that $P^0(\mb{s},\bm{s}')$ is the identity matrix and conditioning the distribution on the initial state $\bm{s}$:
\begin{equation}
 P_{T|S}(t|\mb{s})=\sum_{\bm{s}'\in\mc{S}_{\beta_{s_0}(t)}}\frac{P^{\beta_{s_0}(t)-b}(\bm{s},\bm{s}')}{N_{s_0}\sigma(\mb{s})}\left[N_{s_0}-s'_{s_0}\left(1-P_{D|S}\left(t-\tau(b,\beta_{s_0}(t))\big|\bm{s}'\right)\right)\right].
\end{equation}
The overall latency \gls{cdf} for a frame in batch $b$ can be computed using the law of total probability:
\begin{equation}
 P_T^{(b)}(t)=\sum_{\bm{s}\in\mc{S}_b}\frac{\pi(\bm{s})P_{T|S}(t|\mb{s})}{\sigma_b\sum_{\bm{s}'\in\mc{S}_b}\pi(\bm{s}')}.
\end{equation}
Due to the need to multiply matrix $\mb{P}$ to obtain the $B$-step transition matrix, the computational complexity of computing the latency \gls{cdf} is $O(|\mc{S}|^3)$. In the worst case, i.e., when all clients arrive at different times, this corresponds to $O(N^3 2^{3N})$.

We can now compute the \gls{paoi} for the system with batch arrivals. In this case, we cannot consider the success probability of successive frames to be independent, as the system is stateful: in order to compute the \gls{paoi}, we need to start immediately after a successful transmission. We know that, if we go from state $\bm{s}$ to state $\bm{s}'$, with $s_0=s_0'=b$, the success probability for a frame belonging to batch $b$ is equal to $\frac{N_{s_0}-s_{s_0}'}{N_{s_0}}$. By applying Bayes' theorem, we have:
\begin{equation}
 P(\bm{s},\bm{s}'|D)=\frac{(N_{s_0}-s_{s_0}')P^B(\bm{s},\bm{s}')}{N_{s_0}\sigma(\bm{s})},
\end{equation}
where $D$ indicates that a frame generated in state $s$ was successfully delivered. We removed the exponent from the notation, as we only consider transitions of $B$ steps in this case. We can also consider the transition probability in case of failure, denoted as $\bar{D}$:
\begin{equation}
 P\left(\bm{s},\bm{s}'|\bar{D}\right)=\frac{s_{s_0}'P^B(\bm{s},\bm{s}')}{N_{s_0}(1-\sigma(\bm{s}))}.
\end{equation}
The transition matrix in the case of failures is then denoted as $\mb{P}_{\bar{D}}$, and its elements are simply given by:
\begin{equation}
 P_{\bar{D}}(\bm{s},\bm{s}')=(1-\sigma(\bm{s}))P\left(\bm{s},\bm{s}'|\bar{D}\right).\label{eq:failure_matrix}
\end{equation}
We also define the transition probability from $\mb{s}$ to $\mb{s}'$ over $m\geq1$ steps, where the first frame is the only successful one, which we denote as $P_{\varnothing}\left(\mb{s},\mb{s}';m\right)$:
\begin{equation}
 P_{\varnothing}\left(\mb{s},\mb{s}';m\right)=\sum_{\mb{s}''\in\mc{S}_b} P(\bm{s},\bm{s}''|D)P_{\bar{D}}^{m-1}(\mb{s}'',\mb{s}'),\label{eq:empty_trans}
\end{equation}
where $P_{\bar{D}}^{m-1}(\mb{s},\mb{s}')$ is 1 if $\mb{s}=\mb{s}'$ and 0 otherwise. We can then give the \gls{cdf} of the \gls{paoi}:
\begin{equation}
\begin{aligned}
P_{\Psi}^{(b)}(\psi)=  u(\psi-\tau)\sum\limits_{\mathclap{\mb{s}\in(\mc{S}_b)}}\frac{\pi(\mb{s})\sigma(\mb{s})\left[1-\sum_{\mb{s}'\in\mc{S}_b}P_{\varnothing}\!\left(\mb{s},\mb{s}';\left\lfloor\frac{\psi}{\tau}\right\rfloor\right)\left(1-\sigma(\mb{s}')P_{T|S}(\text{mod}(\psi,\tau)|\mb{s}')\right)\right]}{\sum_{\mb{s^*}\in\mc{S}_b}\pi(\mb{s}^*)\sigma(\mb{s}^*)}.
\label{eq:batch_paoi}
\end{aligned}
\end{equation}
where $\mathbb{N}^+$ is the set of strictly positive integers. In order to compute the latency distribution, we need a further $O\left(\left\lfloor\frac{\psi}{\tau}\right\rfloor |\mc{S}|^3\right)$ operations. The complexity of computing the \gls{paoi} \gls{cdf} is then $O\left(\left(\left\lfloor\frac{\psi}{\tau}\right\rfloor+1\right) |\mc{S}|^3\right)$.

We can finally compute the \gls{aoi} for the system: as for the synchronized case, we can use the geometric method and obtain the formula in~\eqref{eq:trapezoids}. However, computing the values in the formula is significantly harder. The inter-arrival time $Y$ is an integer multiple of the period $\tau$, and its \gls{pmf} $p_Y^{(b)}(m\tau)$ is given by:
\begin{equation}
 p_Y^{(b)}(m\tau)=\sum\limits_{\mathclap{\mb{s}\in(\mc{S}_b)}}\frac{\pi(\mb{s})\sigma(\mb{s})\sum_{\mb{s}'\in\mc{S}_b}P_{\varnothing}(\mb{s},\mb{s}';m)\sigma(\mb{s}')}{\sum_{\mb{s^*}\in\mc{S}_b}\pi(\mb{s}^*)\sigma(\mb{s}^*)}.
\end{equation}
We can expand the transition probability $P_{\varnothing}(\mb{s},\mb{s}';m)$ to obtain:
\begin{equation}
  p_Y^{(b)}(m\tau)=\sum\limits_{\mathclap{\mb{s}\in(\mc{S}_b)}}\frac{\pi(\mb{s})\sigma(\mb{s})}{\sum_{\mb{s^*}\in\mc{S}_b}\pi(\mb{s}^*)\sigma(\mb{s}^*)}\sum_{\mb{s}'\in\mc{S}_b}\sum_{\mb{s}''\in\mc{S}_b}P(\mb{s},\mb{s}''|D)P_{\bar{D}}^{m-1}(\mb{s}'',\mb{s}')\sigma(\mb{s}').
\end{equation}
Assuming that $\mb{P}_{\bar{D}}$ is diagonalizable, i.e., $\mb{P}_{\bar{D}}=\mb{V}\mb{A}\mb{V}^{-1}$, where $\mb{A}$ is a diagonal matrix (and $\mb{d}$ is the vector representing its diagonal), we can transform the \gls{pmf} to the following form:
\begin{equation}
  p_Y^{(b)}(m\tau)=\sum\limits_{{\mb{s}\in(\mc{S}_b)}}\frac{\pi(\mb{s})\sigma(\mb{s})}{\sum_{\mb{s^*}\in\mc{S}_b}\pi(\mb{s}^*)\sigma(\mb{s}^*)}\sum_{\mb{s}'\in\mc{S}_b}\sum_{\mb{s}''\in\mc{S}_b}\sigma(\mb{s}')P(\mb{s},\mb{s}''|D)\left[\mb{V}\mb{A}^{m-1}\mb{V}^{-1}\right](\mb{s}'',\mb{s}').\label{eq:py_diag}
\end{equation}
Since $\mb{P}_{\bar{D}}$ is not diagonalizable in the general case, we adopt a perturbation approach~\cite{greenbaum2020first}, adding a small Gaussian noise (in the simulations, we set the variance to $10^{-6}$) to each component of its diagonal by summing a diagonal noise matrix $\mb{W}$ to $\mb{A}$. This approach~\cite{davies2008approximate}, which approximates $\mb{A}$ as $\mb{V}(\mb{A}+\mb{W})\mb{V}^{-1}+\mb{E}$, guarantees a small value of $||\mb{E}||$ under our conditions. Since we are taking successive powers, the diagonal approximation will have an increasing error, but the probability of having a large number of consecutive failures is relatively small when we operate close to the optimal load, as our numerical results will show. We note that in the synchronized case, i.e., $B=1$, the transition matrix has identical rows, resulting in a large error from the naive approximated diagonalization.

We can then compute $\E{Y}$ as follows:
\begin{equation}
 \begin{aligned}
  \E{Y}=&\sum_{m=1}^{\infty}m\tau p_Y^{(b)}(m\tau)\\
  =&\sum_{m=1}^{\infty}m\tau\sum\limits_{{\mb{s}\in(\mc{S}_b)}}\frac{\pi(\mb{s})\sigma(\mb{s})}{\sum_{\mb{s^*}\in\mc{S}_b}\pi(\mb{s}^*)\sigma(\mb{s}^*)}\sum_{\mb{s}'\in\mc{S}_b}\sum_{\mb{s}''\in\mc{S}_b}\sigma(\mb{s}')P(\mb{s},\mb{s}''|D)\left[\mb{V}\mb{A}^{m-1}\mb{V}^{-1}\right](\mb{s}'',\mb{s}')\\
  =&\sum\limits_{{\mb{s}\in(\mc{S}_b)}}\frac{\pi(\mb{s})\sigma(\mb{s})\tau}{\sum_{\mb{s^*}\in\mc{S}_b}\pi(\mb{s}^*)\sigma(\mb{s}^*)}\sum_{\mb{s}'\in\mc{S}_b}\sum_{\mb{s}''\in\mc{S}_b}\sigma(\mb{s}')P(\mb{s},\mb{s}''|D)\left[\mb{V}\left(\sum_{m=1}^{\infty}m\mb{A}^{m-1}\right)\mb{V}^{-1}\right](\mb{s}'',\mb{s}').
 \end{aligned}
\end{equation}
The result of the infinite series is simple, as each element of the diagonal matrix $\mb{A}$ can be computed independently. We can then compute the matrix $\mb{Z}$ that solves the series:
\begin{equation}
 Z(i,j)=\begin{cases}
         \frac{1}{(1-A(i,i))^2}, &\text{if }i=j;\\
         0, &\text{otherwise.}
        \end{cases}
\end{equation}
The expected time between two consecutive successful frames is then:
\begin{equation}
 \E{Y}=\sum\limits_{{\mb{s}\in(\mc{S}_b)}}\frac{\pi(\mb{s})\sigma(\mb{s})\tau}{\sum_{\mb{s^*}\in\mc{S}_b}\pi(\mb{s}^*)\sigma(\mb{s}^*)}\sum_{\mb{s}'\in\mc{S}_b}\sum_{\mb{s}''\in\mc{S}_b}\sigma(\mb{s}')P(\mb{s},\mb{s}''|D)\left[\mb{V}\mb{U}\mb{V}^{-1}\right](\mb{s}'',\mb{s}')\label{eq:y_batch}
\end{equation}
We can compute $\E{Y^2}$ in the same way, defining matrix $\mb{U}$ as follows:
\begin{align}
U(i,j)=&\begin{cases}
         \frac{1+A(i,i)}{(1-A(i,i))^3}, &\text{if }i=j;\\
         0, &\text{otherwise.}
        \end{cases}\\
 \E{Y^2}=&\sum\limits_{{\mb{s}\in(\mc{S}_b)}}\frac{\pi(\mb{s})\sigma(\mb{s})\tau^2}{\sum_{\mb{s^*}\in\mc{S}_b}\pi(\mb{s}^*)\sigma(\mb{s}^*)}\sum_{\mb{s}'\in\mc{S}_b}\sum_{\mb{s}''\in\mc{S}_b}\sigma(\mb{s}')P(\mb{s},\mb{s}''|D)\left[\mb{V}\mb{W}\mb{V}^{-1}\right](\mb{s}'',\mb{s}')\label{eq:y2_batch}
\end{align}
We can then compute $\E{TY}$, knowing that $T$ only depends on the state when the frame is generated. We can then apply the law of total probability, knowing that the final state $\mb{s}'$ is part of the calculation, and get:
\begin{equation}
\begin{aligned}
 \E{TY}=&\sum\limits_{{\mb{s}\in(\mc{S}_b)}}\frac{\pi(\mb{s})\sigma(\mb{s})\tau}{\sum_{\mb{s^*}\in\mc{S}_b}\pi(\mb{s}^*)\sigma(\mb{s}^*)}\sum_{\mb{s}'\in\mc{S}_b}\sum_{\mb{s}''\in\mc{S}_b}\sigma(\mb{s}')\E{T|S=\mb{s'}}P(\mb{s},\mb{s}''|D)\left[\mb{V}\mb{Z}\mb{V}^{-1}\right](\mb{s}'',\mb{s}').\label{eq:ty_batch}
\end{aligned}
\end{equation}
We can then finally compute the expected latency, dividing each segment:
\begin{equation}
 \begin{aligned}
  \E{T|S=\mb{s}}=&\int_0^{\tau}1-P_{T|S}(t|\mb{s})dt\\
  =&\tau-\int\displaylimits_0^{\mathclap{\nu_{s_0+1}}}\frac{P_{D|S}(t|\mb{s})}{\sigma(\mb{s})}dt+
  \sum_{b=s_0+1}^{B+s_0-1}\int\displaylimits_0^{\mathclap{\nu_{b+1}}}\sum_{\mb{s}'\in\mc{S}_b}\frac{P^{b-s_0}(\mb{s},\mb{s}')}{\sigma(\mb{s})N_{s_0}}\left(N_{s_0}-s_0'\left(1-P_{D|S}(t|\mb{s}')\right)\right)dt\\
  =&-\frac{(1-\sigma(\mb{s}))\tau}{\sigma(\mb{s})}+\sum_{k=0}^{X(\mb{s})-1}\frac{(X(\mb{s})-k)\gamma_{k+1}(\mu\nu_{s_0+1})}{\mu X(\mb{s})\sigma(\mb{s})k!}+\sum_{b=s_0+1}^{B+s_0-1}\sum_{\mb{s}'\in\mc{S}_b}\frac{s'_{s_0}P^{b-s_0}(\mb{s},\mb{s}')}{\mu\sigma(\mb{s})N_{s_0}X(\mb{s}')}\\
  &\times\sum_{k=0}^{X(\mb{s}')-1}\frac{(X(\mb{s}')-k)\gamma_{k+1}(\mu\nu_{b+1})}{k!}.
 \end{aligned}\label{eq:et_state}
\end{equation}
The expected \gls{aoi} can then be approximated by plugging the results from~\eqref{eq:y_batch}-\eqref{eq:et_state} into~\eqref{eq:trapezoids}.

\subsection{FIFO Queuing}

If we consider a \gls{fifo} queue, the state definition can be significantly simplified, as the order of service is fixed: frames from older batches are always served first, with the full rate $\mu$ of the \gls{mec} server. We can then reduce the state to only 2 values:
\begin{itemize}
 \item The index $s_0$ of the next batch;
 \item The number $s_1\in\{0,\ldots,N\}$ of frames (from any batch) that are still in processing immediately before the next batch is generated, i.e., at time $\nu_{s_0}-\varepsilon$.
\end{itemize}
The state vector is then $\mb{s}=(s_0,s_1)$, and the state space of the Markov chain representing the system is $\mc{S}=\{0,\ldots,B\}\times\{0,\ldots,N\}$. The state space size is $B(N+1)$, and in the most extreme case, each frame arrives at a different time and the state space size is $N(N+1)$. The subset $\mc{S}_b$ has the same meaning as for the \gls{gps} case, as defined by~\eqref{eq:sb}. We can now define the set of reachable states from state $\mb{s}$, $\mc{R}(\mb{s})$:
\begin{equation}
 \mc{R}(\mb{s})=\left\{\mb{s}'\in\mc{S}_{s_0+1}:s_1'\leq\min\left(\left\{s_1+N_{s_0},N\right\}\right)\right\}.
\end{equation}
The minimum in the set is due to preemption: if any frames from batch $s_0$ are still in the system when new frames are generated, they are superseded by their newer version and simply discarded. We can also define $X(\mb{s})$ in a similar way to the version we gave for the \gls{gps} system:
\begin{equation}
 X(\mb{s})=\min\left(\left\{s_1+N_{s_0},N\right\}\right).
\end{equation}
The definition of the \gls{pmf} of the number of events in interval $\nu(\mb{s})$ is then the same as the one given in~\eqref{eq:events_state} for the \gls{gps} case. The state transition is then simple:
\begin{equation}
 P(\bm{s},\bm{s}')=\begin{cases}
  p_{C|S,T}\left(X(\bm{s})-s_1'|\bm{s},\nu_{s_0+1}\right), &\text{if }\bm{s}'\in\mc{R}(s);\\
  0, &\text{otherwise.}
 \end{cases}
\end{equation}
However, frames in a \gls{fifo} $D^{N_b}/M/1$ queuing system are served one at a time in strict order, instead of sharing the processor. If we consider a frame of batch $b$, the frames in the batch will be at the head of the queue, so that the success probability is simple. If the length of the queue after a full period $\tau$ (i.e., after $B$ steps of the Markov chain) is lower than $N-N_b$, all frames from the batch have been served. Otherwise, only a few (or none) have, and the selection of the served frames within the batch is random:
\begin{equation}
\sigma(\mb{s})=\sum_{s_1'=0}^{N-N_{s_0}}P^B(\mb{s},(s_0,s_1'))+\sum_{s_1'=N-N_{s_0}+1}^{N-1}P^B(\mb{s},(s_0,s_1'))\frac{N-s_1'}{N_{s_0}}.\label{eq:fifo_state_succ}
\end{equation}
As for the \gls{gps} case, the stationary state distribution $\bm{\pi}$ can be computed using the eigenvector method. We can then follow the same procedure as we did for the \gls{gps} case to derive the latency and \gls{paoi} distributions, after defining the number $Q_b(b')$ of frames behind the ones generated in batch $b$ when the last generated batch is $b'$:
\begin{equation}
Q_b(b')=\begin{cases}
        \sum_{i=b+1}^{b'}N_i,                         &\text{if }b<b';\\
        \sum_{i=b+1}^{B}N_i+\sum_{i=1}^{b'}N_i,       &\text{if }b>b';\\
        0,                                            &\text{if }b=b'.
        \end{cases}
\end{equation}
The success probability for a frame in batch $b$ follows~\eqref{eq:tot_success}, using the state success probability from~\eqref{eq:fifo_state_succ}.  The probability of a frame from batch $b$ being served by time $t$, starting from state $\mb{s}$ at time 0 and with $t\leq\nu_{s_0}$, i.e., with no new frames generated between $0$ and $t$, is:
\begin{equation}
 P_{D|S}^{(b)}(t|\mb{s})=\begin{cases}
0, &\text{if }X(\mb{s})\leq Q_b(s_0);\\
\quad\ \ \sum\limits_{\mathclap{c=X(\mb{s})-Q_b(s_0)}}^{X(\mb{s})} p_{C|S,T}(c|\mb{s},t)+\sum\limits_{\mathclap{c=[X(\mb{s})-Q_b(s_0)-N_b]^{+}+1}}^{\mathclap{X(\mb{s})-Q_b(s_0)-1}}p_{C|S,T}(c|\mb{s},t)\frac{(c-[X(\mb{s})-N_b-Q_b(s_0]^{+})}{\min\left(\left\{N_b,X(\mb{s})-Q_b(s_0)\right\}\right)}, &\text{otherwise,}
\end{cases}
\end{equation}
where $[x]^+$ denotes the positive part function, equal to $x$ if $x\geq0$ and 0 otherwise.
If $X(\mb{s})\leq Q_b(s_0)$, all frames from the batch have already been served. Using the definition of $\beta_b(t)$ from~\eqref{eq:batch_idx} and the definition of $\tau(b,b')$ from~\eqref{eq:batch_time}, we can then give the \gls{cdf} of the latency in the case of a transition from $\mb{s}$ to $\mb{s'}$:
\begin{equation}
 P_{T|S}(t|\mb{s},\mb{s}')=\begin{cases}
                    1, &\text{if }X(\mb{s}')\leq Q_{s_0}(\beta_{s_0}(t));\\
                    1-\frac{\left(1-P^{(s_0)}_{D|S}\left(t-\tau(s_0,\beta_{s_0}(t))|\bm{s}'\right)\right)}{N_{s_0}(X(\bm{s}')-Q_{s_0}(\beta_{s_0}(t)))^{-1}}, &\text{if }Q_{s_0}(\beta_{s_0}(t))<X(\mb{s}')<Q_{s_0}(\beta_{s_0}(t))+N_{s_0};\\
                    P^{(s_0)}_{D|S}\left(t-\tau(s_0,\beta_{s_0}(t))|\bm{s}'\right), &\text{if }X(\mb{s}')\geq Q_{s_0}(\beta_{s_0}(t))+N_{s_0}.
                   \end{cases}
\end{equation}
We then apply the law of total probability to remove the condition on $\mb{s}'$:
\begin{equation}
 P_{T|S}(t|\mb{s})=\left(\sigma(\mb{s})\right)^{-1}\sum_{s_1'=0}^{N}P^{\beta_{s_0}(t)-b}(\bm{s},(\beta_{s_0}(t),s_1'))P_{T|S}(t|\mb{s},\mb{s}').
\end{equation}
We can then get the \gls{cdf} of the latency by applying the law of total probability a second time:
\begin{equation}
 P_T^{(b)}(t)=\sum_{\mb{s}\in\mc{S}_b}\frac{\pi(\bm{s})\sigma(\mb{s})P_{T|S}(t|\mb{s})}{\sum_{\mb{s}'\in\mc{S}_b}\pi(\mb{s}')\sigma(\mb{s})}.
\end{equation}

Following the same procedure that we used for \gls{gps} service, we can give the transition probability in case of success. If we go from state $\bm{s}$ to state $\bm{s}'$, with $s_0=s'_0=b$, the success probability for a frame belonging to batch $b$ is equal to $\min\left(\left\{1,\frac{(N-s_1')}{N_b}\right\}\right)$. The conditional transition probability, $P(\bm{s},\bm{s}'|D)$, can then be computed by applying Bayes' theorem:
\begin{equation}
 P(\bm{s},\bm{s}'|D)=\begin{cases}
    \frac{P^B(\bm{s},\bm{s}')}{\sigma(\bm{s})}, &\text{if }s_1'\leq N-N_{s_0};\\
    \frac{(N-s_1')P^B(\bm{s},\bm{s}')}{N_{s_0}\sigma(\bm{s})}, &\text{otherwise.}
                     \end{cases}
\end{equation}
We also compute the conditional transition probability in case of failure, $P\left(\bm{s},\bm{s}'|\bar{D}\right)$:
\begin{equation}
 P\left(\bm{s},\bm{s}'|\bar{D}\right)=\begin{cases}
    0, &\text{if }s_1'\leq N-N_{s_0};\\
    \frac{(s_1'-N_{s_0})P^B(\bm{s},\bm{s}')}{N_{s_0}(1-\sigma(\bm{s}))}, &\text{otherwise.}
                     \end{cases}
\end{equation}
The rest of the derivation of the \gls{paoi} \gls{cdf} is the same: the transition matrix in case of a failure $\mb{P}_{\bar{D}}$ is computed as in~\eqref{eq:failure_matrix}. The transition probability over $m$ steps, where the first frame is the only successful one, is given by~\eqref{eq:empty_trans}, and finally, the \gls{paoi} \gls{cdf} can be obtained by using the \gls{fifo} versions of the terms in~\eqref{eq:batch_paoi}. The complexity of the latency and \gls{paoi} calculations is the same as the \gls{gps} case, but we note that, as the state space is significantly smaller for most practical cases, the complexity will be correspondingly lower.

In order to derive the expected \gls{aoi}, we can directly apply~\eqref{eq:y_batch}-\eqref{eq:ty_batch} and plug them into~\eqref{eq:trapezoids}, while the value of $\E{T|S=\mb{s}}$ is given by:
\begin{equation}
 \begin{aligned}
  \E{T|S=\mb{s}}=&\int_0^{\tau}1-P_{T|S}(t|\mb{s})dt\\
  =&\tau-\int\displaylimits_0^{\mathclap{\nu_{s_0+1}}}\frac{P_{D|S}(t|\mb{s})}{\sigma(\mb{s})}dt+
  \sum_{b=s_0+1}^{B+s_0-1}\int\displaylimits_0^{\mathclap{\nu_{b+1}}}\sum_{\mb{s}'\in\mc{S}_b}\frac{P^{b-s_0}(\mb{s},\mb{s}')}{\sigma(\mb{s})N_{s_0}}\left(N_{s_0}-s_0'\left(1-P_{D|S}(t|\mb{s}')\right)\right)dt\\
  =&-\frac{(1-\sigma(\mb{s}))\tau}{\sigma(\mb{s})}+\sum_{b=s_0}^{B+s_0-1}\sum_{\mb{s}'\in\mc{S}_b}\frac{P^{b-s_0}(\mb{s},\mb{s}')\min\left(1,\frac{X(\mb{s}')-Q_{s_0}(b)}{N_{s_0}}\right)}{\mu\sigma(\mb{s})}\\
  &\times\sum_{k=0}^{X(\mb{s}')-Q_{s_0}(b)-1}\frac{\min\left(1,\frac{X(\mb{s}')-Q_{s_0}(b)-k}{N_{s_0}}\right)\gamma_{k+1}(\mu\nu_{b+1})}{k!}.
 \end{aligned}\label{eq:et_state_dm1}
\end{equation}

\section{Simulation Settings and Results}\label{sec:results}

We now verify the correctness of the analysis by Monte Carlo simulation, considering $L=10^7$ cycles for each scenario and investigating the main design parameters and choices when dimensioning an edge computing server, namely, the update frequency, the computational power of the server, and the queuing policy and scheduling used to control requests. The parameters for the simulations are specified for each scenario. The units given here are dimensionless, as they are all relative (i.e., if the time unit is 1~ms, the period $\tau$ is expressed in ms and the service rate $\mu$ is expressed in frames/ms). Realistic values can be considered for each target application, with no change in the main design considerations.

\subsection{Synchronized Frame Generation}

\begin{figure*}[t!]
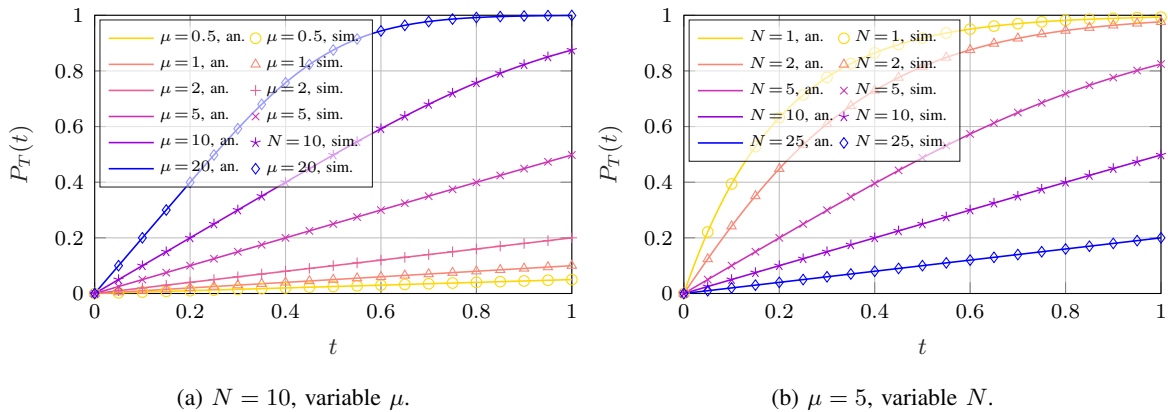

    \centering
\subfloat[$N=10$, variable $\mu$.\label{fig:sync_mu_latency}]
{\input{./fig/sync_mu_lat_cdf.tex}}
\subfloat[$\mu=5$, variable $N$.\label{fig:sync_N_latency}]
{\input{./fig/sync_N_lat_cdf.tex}}
    \caption{Latency \gls{cdf} for synchronized frame generation (frames with a latency larger than $\tau=1$ are dropped due to preemption).}
    \label{fig:sync_lat}
\end{figure*}

Firstly, we consider the synchronized frame generation scenario, in which all frames arrive at the same time. In this case, the \gls{gps} and \gls{fifo} policies have identical results. Fig.~\ref{fig:sync_lat} shows the latency \glspl{cdf} for different systems, considering $\tau=1$: firstly, we can note that the Monte Carlo results match the theoretical curves perfectly, confirming the correctness of the derivation. A system with $N=10$ and different values of the server computational power $\mu$ is shown in Fig.~\ref{fig:sync_mu_latency}, while Fig.~\ref{fig:sync_N_latency} shows a system with $\mu=5$ and a variable number of clients $N$. As expected, a lower number of clients, or a larger computational power, lead to a lower latency and a higher reliability: most of the \glspl{cdf} do not reach 1, as frames which are not processed after an interval $\tau$ are discarded and superseded by newer ones from the same clients.

We can also note that the system should be overdimensioned: if $\mu\tau=N^{-1}$, i.e., the average load on the server is 1, between 10 and 20\% of frames are discarded. We also note that the latency distribution tends to the uniform distribution as the load becomes higher, as would be expected from Poisson departures, while systems with a lighter load have a linear \gls{cdf} in the first part of the interval, then a lower probability in the second, as the probability of having already served the relevant frame is significantly higher.

\begin{figure*}[t!]
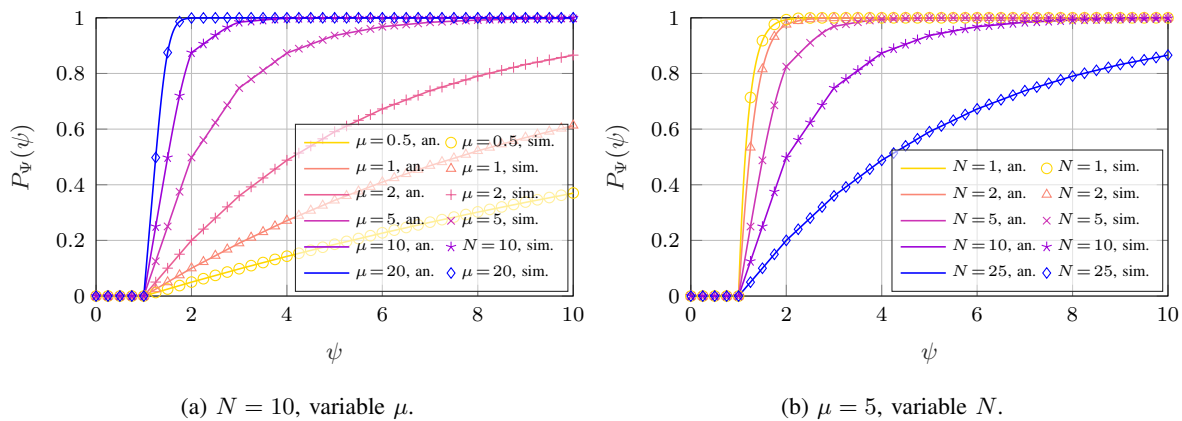

    \centering
\subfloat[$N=10$, variable $\mu$.\label{fig:sync_mu_paoi}]
{\input{./fig/sync_mu_paoi_cdf.tex}}
\subfloat[$\mu=5$, variable $N$.\label{fig:sync_N_paoi}]
{\input{./fig/sync_N_paoi_cdf.tex}}
    \caption{\gls{paoi} \gls{cdf} for synchronized frame generation with $\tau=1$.}
    \label{fig:sync_paoi}
\end{figure*}

We can also look at the \gls{paoi} distribution in the same scenarios, shown in Fig.~\ref{fig:sync_paoi}: in this case, we can easily see that even relatively underdimensioned systems still have a good worst-case performance. If we consider a system with $N=10$ clients, as shown in Fig.~\ref{fig:sync_mu_paoi}, a system with $\mu=10$ can guarantee a \gls{paoi} lower than 4 with a probability higher than 99\%. Even an underdimensioned system with $\mu=5$ has a \gls{paoi} lower than 4 about 90\% of the time. This is due to the resistance of \gls{paoi} to discarded frames: even if one or two consecutive frames are lost, a preemptive system which avoids queuing can still deliver fresh information.

\begin{figure*}[t!]
    \centering
\subfloat[$N=10$, variable $\mu$.\label{fig:sync_mu_aoi}]
{\input{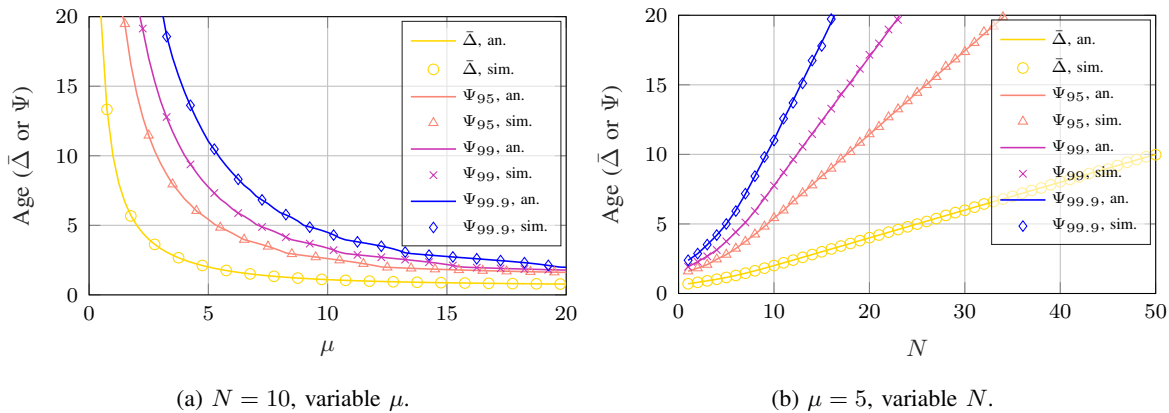}}
\subfloat[$\mu=5$, variable $N$.\label{fig:sync_N_aoi}]
{\begin{tikzpicture}

\begin{axis}[%
width=\sfwidth,
height=\sfheight,
xmin=0,
xmax=50,
legend style={legend cell align=left, fill opacity=0.6, draw opacity=1, text opacity=1, legend columns=1, align=left, draw=white!15!black, font=\tiny, at={(0.99, 0.985)}, anchor=north east},
xlabel style={font=\footnotesize\color{white!15!black}},
xlabel={$N$},
ymin=0,
ymax=20,
xmajorgrids,
ymajorgrids,
ylabel style={font=\footnotesize\color{white!15!black}},
ylabel={Age ($\bar{\Delta}$ or $\Psi$)},
axis background/.style={fill=white}
]
\addplot [color=color0, semithick]
  table[row sep=crcr]{%
1	0.7\\
2	0.801725171140759\\
3	0.908338505642713\\
4	1.02411031504677\\
5	1.15298074906399\\
6	1.29733689427626\\
7	1.45737079914467\\
8	1.63119429092703\\
9	1.81562062034554\\
10	2.00720856196781\\
11	2.20307321299859\\
12	2.4012146803402\\
13	2.60044686957686\\
14	2.80015363249569\\
15	3.00004954685121\\
16	3.20001504185918\\
17	3.40000431250025\\
18	3.60000117101822\\
19	3.80000030196512\\
20	4.00000007412413\\
21	4.20000001735938\\
22	4.40000000388654\\
23	4.60000000083341\\
24	4.80000000017146\\
25	5.0000000000339\\
26	5.20000000000645\\
27	5.40000000000118\\
28	5.60000000000021\\
29	5.80000000000003\\
30	6.00000000000001\\
31	6.2\\
32	6.4\\
33	6.6\\
34	6.8\\
35	7\\
36	7.2\\
37	7.4\\
38	7.6\\
39	7.8\\
40	8\\
41	8.2\\
42	8.4\\
43	8.6\\
44	8.8\\
45	9\\
46	9.2\\
47	9.39999999999999\\
48	9.6\\
49	9.8\\
50	10\\
};
\addlegendentry{$\bar{\Delta}$, an.}

\addplot [only marks, mark=o, mark options={solid, draw=color0}]
  table[row sep=crcr]{%
1	0.700207740892698\\
2	0.802191719679656\\
3	0.908105393460163\\
4	1.02403880983474\\
5	1.15421527989007\\
6	1.29668299644296\\
7	1.45746793108684\\
8	1.63067022586133\\
9	1.82114450270228\\
10	2.00831264467239\\
11	2.20677334014313\\
12	2.40002923259914\\
13	2.60046167481089\\
14	2.79790060926452\\
15	2.99841054239137\\
16	3.20192190308116\\
17	3.41019380914473\\
18	3.59036584975208\\
19	3.79982206177117\\
20	4.01036629026091\\
21	4.20087501112082\\
22	4.4104543417734\\
23	4.592738675205\\
24	4.79380222526569\\
25	4.98870048211988\\
26	5.17750466978798\\
27	5.3847637687235\\
28	5.61394552676695\\
29	5.81444361328304\\
30	5.98267635828846\\
31	6.20536631773053\\
32	6.43490771476655\\
33	6.61005147962834\\
34	6.78805233596886\\
35	6.98126558322183\\
36	7.20518777077398\\
37	7.40516804774998\\
38	7.58314604813415\\
39	7.81890025513663\\
40	8.04324873656944\\
41	8.2018182001105\\
42	8.38005195748849\\
43	8.62290990621415\\
44	8.79756268753407\\
45	8.93938128967793\\
46	9.19641256626757\\
47	9.4291086667368\\
48	9.63361475988165\\
49	9.83242416675521\\
50	9.97736241350024\\
};
\addlegendentry{$\bar{\Delta}$, sim.}

\addplot [color=color2, semithick]
  table[row sep=crcr]{%
1	1.6\\
2	1.823\\
3	2.077\\
4	2.459\\
5	2.796\\
6	3.232\\
7	3.737\\
8	4.255\\
9	4.819\\
10	5.429\\
11	5.966\\
12	6.628\\
13	7.209\\
14	7.817\\
15	8.438\\
16	8.997\\
17	9.643\\
18	10.234\\
19	10.833\\
20	11.449\\
21	12.019\\
22	12.65\\
23	13.243\\
24	13.84\\
25	14.453\\
26	15.03\\
27	15.652\\
28	16.247\\
29	16.844\\
30	17.454\\
31	18.035\\
32	18.653\\
33	19.248\\
34	19.845\\
35	21\\
36	21\\
37	21\\
38	21\\
39	21\\
40	21\\
41	21\\
42	21\\
43	21\\
44	21\\
45	21\\
46	21\\
47	21\\
48	21\\
49	21\\
50	21\\
};
\addlegendentry{$\Psi_{95}$, an.}

\addplot [only marks, mark=triangle, mark options={solid, draw=color2}]
  table[row sep=crcr]{%
1	1.599\\
2	1.823\\
3	2.075\\
4	2.462\\
5	2.798\\
6	3.233\\
7	3.732\\
8	4.252\\
9	4.821\\
10	5.433\\
11	5.978\\
12	6.634\\
13	7.214\\
14	7.784\\
15	8.449\\
16	8.986\\
17	9.647\\
18	10.179\\
19	10.848\\
20	11.461\\
21	11.984\\
22	12.678\\
23	13.251\\
24	13.834\\
25	14.437\\
26	14.995\\
27	15.579\\
28	16.309\\
29	16.839\\
30	17.368\\
31	18.02\\
32	18.795\\
33	19.196\\
34	19.867\\
35	21\\
36	21\\
37	21\\
38	21\\
39	21\\
40	21\\
41	21\\
42	21\\
43	21\\
44	21\\
45	21\\
46	21\\
47	21\\
48	21\\
49	21\\
50	21\\
};
\addlegendentry{$\Psi_{95}$, sim.}

\addplot [color=color4, semithick]
  table[row sep=crcr]{%
1	1.922\\
2	2.277\\
3	2.678\\
4	3.13\\
5	3.736\\
6	4.424\\
7	5.102\\
8	5.927\\
9	6.833\\
10	7.755\\
11	8.684\\
12	9.617\\
13	10.549\\
14	11.479\\
15	12.406\\
16	13.331\\
17	14.253\\
18	15.174\\
19	16.092\\
20	17.01\\
21	17.943\\
22	18.877\\
23	19.808\\
24	21\\
25	21\\
26	21\\
27	21\\
28	21\\
29	21\\
30	21\\
31	21\\
32	21\\
33	21\\
34	21\\
35	21\\
36	21\\
37	21\\
38	21\\
39	21\\
40	21\\
41	21\\
42	21\\
43	21\\
44	21\\
45	21\\
46	21\\
47	21\\
48	21\\
49	21\\
50	21\\
};
\addlegendentry{$\Psi_{99}$, an.}

\addplot [only marks, mark=x, mark options={solid, draw=color4}]
table[row sep=crcr]{%
1	1.921\\
2	2.277\\
3	2.676\\
4	3.13\\
5	3.751\\
6	4.407\\
7	5.105\\
8	5.922\\
9	6.883\\
10	7.754\\
11	8.702\\
12	9.635\\
13	10.552\\
14	11.454\\
15	12.379\\
16	13.287\\
17	14.357\\
18	15.109\\
19	16.152\\
20	17.159\\
21	17.979\\
22	18.979\\
23	19.662\\
24	21\\
25	21\\
26	21\\
27	21\\
28	21\\
29	21\\
30	21\\
31	21\\
32	21\\
33	21\\
34	21\\
35	21\\
36	21\\
37	21\\
38	21\\
39	21\\
40	21\\
41	21\\
42	21\\
43	21\\
44	21\\
45	21\\
46	21\\
47	21\\
48	21\\
49	21\\
50	21\\
};
\addlegendentry{$\Psi_{99}$, sim.}

\addplot [color=color6, semithick]
  table[row sep=crcr]{%
1	2.382\\
2	2.862\\
3	3.5\\
4	4.187\\
5	4.979\\
6	5.978\\
7	7.149\\
8	8.44\\
9	9.74\\
10	11.041\\
11	12.497\\
12	13.863\\
13	15.277\\
14	16.686\\
15	18.045\\
16	19.483\\
17	21\\
18	21\\
19	21\\
20	21\\
21	21\\
22	21\\
23	21\\
24	21\\
25	21\\
26	21\\
27	21\\
28	21\\
29	21\\
30	21\\
31	21\\
32	21\\
33	21\\
34	21\\
35	21\\
36	21\\
37	21\\
38	21\\
39	21\\
40	21\\
41	21\\
42	21\\
43	21\\
44	21\\
45	21\\
46	21\\
47	21\\
48	21\\
49	21\\
50	21\\
};
\addlegendentry{$\Psi_{99.9}$, an.}

\addplot [only marks, mark=diamond, mark options={solid, draw=color6}]
table[row sep=crcr]{%
1	2.389\\
2	2.869\\
3	3.515\\
4	4.198\\
5	4.999\\
6	5.927\\
7	7.176\\
8	8.436\\
9	9.814\\
10	11.001\\
11	12.569\\
12	13.705\\
13	15.107\\
14	16.746\\
15	17.795\\
16	19.727\\
17	21\\
18	21\\
19	21\\
20	21\\
21	21\\
22	21\\
23	21\\
24	21\\
25	21\\
26	21\\
27	21\\
28	21\\
29	21\\
30	21\\
31	21\\
32	21\\
33	21\\
34	21\\
35	21\\
36	21\\
37	21\\
38	21\\
39	21\\
40	21\\
41	21\\
42	21\\
43	21\\
44	21\\
45	21\\
46	21\\
47	21\\
48	21\\
49	21\\
50	21\\
};
\addlegendentry{$\Psi_{99.9}$, sim.}

\end{axis}
\end{tikzpicture}%
}
    \caption{Average \gls{aoi} and \gls{paoi} percentiles for synchronized frame generation with $\tau=1$ as a function of the number of clients $N$ (with $\mu=5$) and of the server rate $\mu$ (with $N=10$ clients).}
    \label{fig:sync_aoi}
\end{figure*}

We can also consider performance in terms of the average \gls{aoi}, or the higher percentiles of the \gls{paoi} distribution (which we denote for brevity's sake as $\Psi_p$, where $p$ is the percentile), shown in Fig.~\ref{fig:sync_aoi}. We can easily see how optimizing the system for the average \gls{aoi} is much less demanding than worst-case optimization: for a scenario with $N=10$ clients, as shown in Fig.~\ref{fig:sync_mu_aoi}, setting $\mu=2$ is sufficient to maintain an \gls{aoi} below 5, while the 99.9th percentile of the \gls{paoi} reaches the same value only if $\mu=9$. The same result is clearly visible as $N$ increases: there is an approximately linear effect for large values of $N$, but the slope for higher percentiles is much steeper, as expected.

\begin{figure*}[t!]
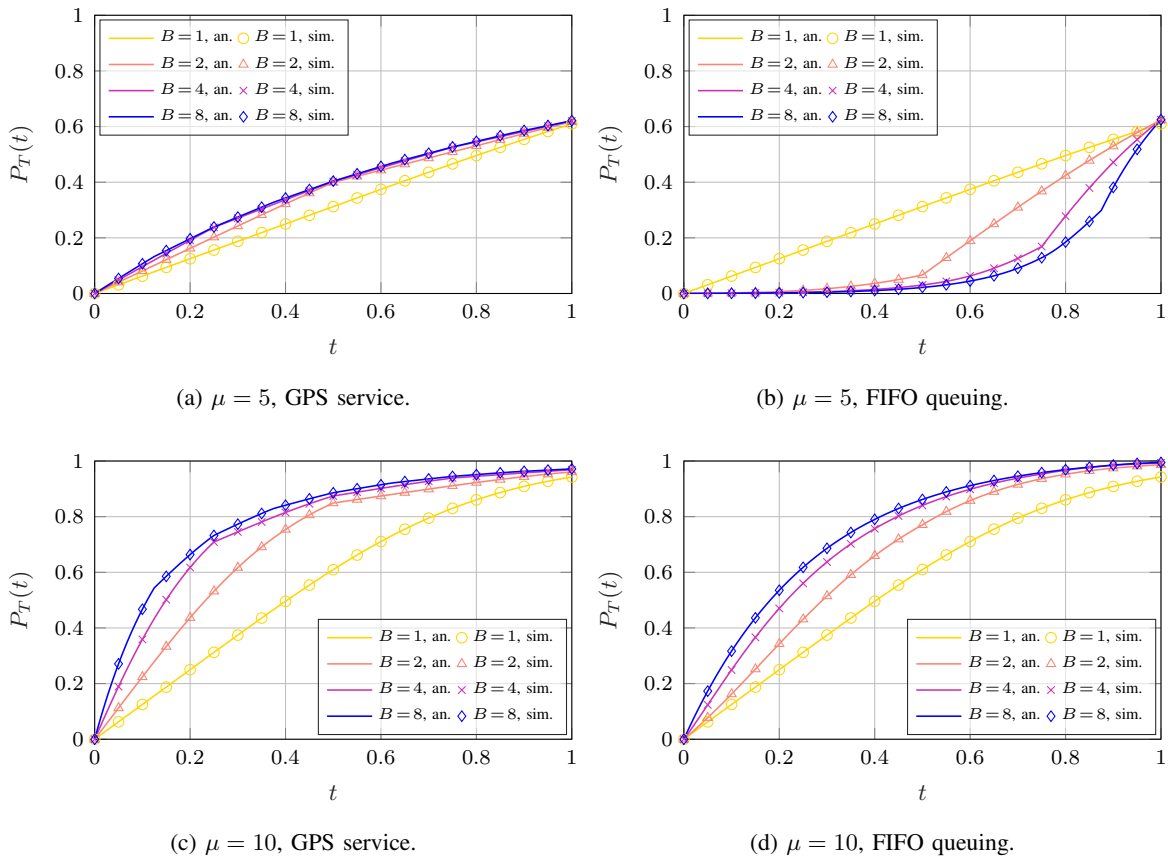

    \centering
\subfloat[$\mu=5$, \gls{gps} service.\label{fig:batGPS_Bh_lat}]
{\input{./fig/batch_Bh_lat_cdf.tex}}
\subfloat[$\mu=5$, \gls{fifo} queuing.\label{fig:batDM1_Bh_lat}]
{\input{./fig/dm1_Bh_lat_cdf.tex}}\\
\subfloat[$\mu=10$, \gls{gps} service.\label{fig:batGPS_Bl_lat}]
{\input{./fig/batch_Bm_lat_cdf.tex}}
\subfloat[$\mu=10$, \gls{fifo} queuing.\label{fig:batDM1_Bl_lat}]
{\input{./fig/dm1_Bm_lat_cdf.tex}}
    \caption{Latency \gls{cdf} for different numbers of identical  batches with $\tau=1$ and $N=8$.}
    \label{fig:batch_B_lat}
\end{figure*}

\subsection{Batch Frame Generation}

We can now consider the more general case of batch frame generation, in which synchronization is not assumed. We first consider the latency \gls{cdf} as a function of the number of batches, fixing $N=8$ and $\tau=1$. Fig.~\ref{fig:batch_B_lat} shows the effect of setting different number of (equally sized) batches on the latency \gls{cdf}: each batch contains $\frac{N}{B}$ clients, and frames are generated at identical intervals $\nu_b=\frac{\tau}{B}$. We can note an interesting difference between the \gls{gps} and \gls{fifo} systems when $\mu=5$, i.e., when the load is relatively high: in the former, whose latency shown in Fig.~\ref{fig:batGPS_Bh_lat}, smaller, more spread out batches improve the latency distribution, as each client has a higher chance of having the whole computational power of the server for itself. On the other hand, the \gls{fifo} system has a significantly worse performance if we increase the number of batches, as frames are always queued, and their latency distribution shifts toward the right. In the \gls{fifo} case, the best latency performance is achieved with $B=1$, which corresponds to the worst case for \gls{gps}. However, in both cases, the overall success probability is similar. If we look at the case with $\mu=10$, i.e., reduce the load by half, the two distributions are much more similar: as shown in Fig.~\ref{fig:batGPS_Bl_lat}-\subref*{fig:batDM1_Bl_lat}, the distribution for the \gls{fifo} system is shifted to the right, but the probability of success is higher, as each frame gets a chance to have the server for itself close to the end of the frame period. In this case, using more batches is beneficial for both systems.

\begin{figure*}[t!]
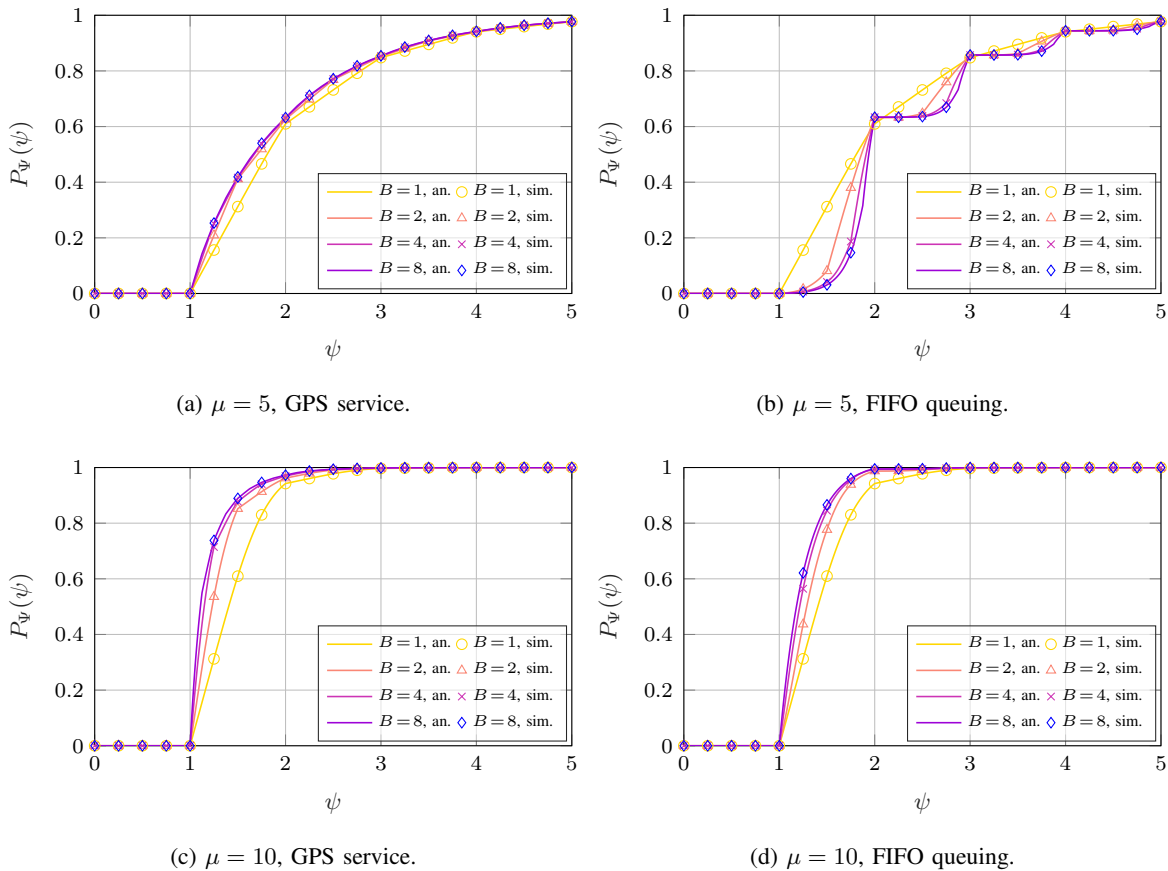

    \centering
\subfloat[$\mu=5$, \gls{gps} service.\label{fig:batGPS_Bh_paoi}]
{\input{./fig/batch_Bh_paoi_cdf.tex}}
\subfloat[$\mu=5$, \gls{fifo} queuing.\label{fig:batDM1_Bh_paoi}]
{\input{./fig/dm1_Bh_paoi_cdf.tex}}\\
\subfloat[$\mu=10$, \gls{gps} service.\label{fig:batGPS_Bl_paoi}]
{\input{./fig/batch_Bm_paoi_cdf.tex}}
\subfloat[$\mu=10$, \gls{fifo} queuing.\label{fig:batDM1_Bl_paoi}]
{\input{./fig/dm1_Bm_paoi_cdf.tex}}
    \caption{\gls{paoi} \gls{cdf} for different numbers of identical batches with $\tau=1$ and $N=8$.}
    \label{fig:batch_Bh_paoi}
\end{figure*}

We can also consider the effect of the batch size on the \gls{paoi}, as we show in Fig.~\ref{fig:batch_Bh_paoi}: we can see an interesting effect in the case with $\mu=5$, as the \gls{gps} system shown in Fig.~\ref{fig:batGPS_Bh_paoi} generally performs better, following a mostly concave curve, while the \gls{cdf} for the \gls{fifo} system shown in Fig.~\ref{fig:batDM1_Bh_paoi} has some clear corner points: whenever a period $\tau$ ends, and the old version of the frame from the client we are considering at the front of the queue is replaced by a new one at the end of the queue, the \gls{pdf} has a discontinuity. As before, the performance of the synchronized system (i.e., $B=1$) is the worst case for \gls{gps} and the best case for \gls{fifo}. On the other hand, if we set $\mu=10$, as in Fig.~\ref{fig:batGPS_Bl_paoi}-\subref*{fig:batDM1_Bl_paoi}, the \gls{fifo} system manages to have a better worst-case performance, and setting $B=8$ is beneficial in both cases.

\begin{figure*}[t!]
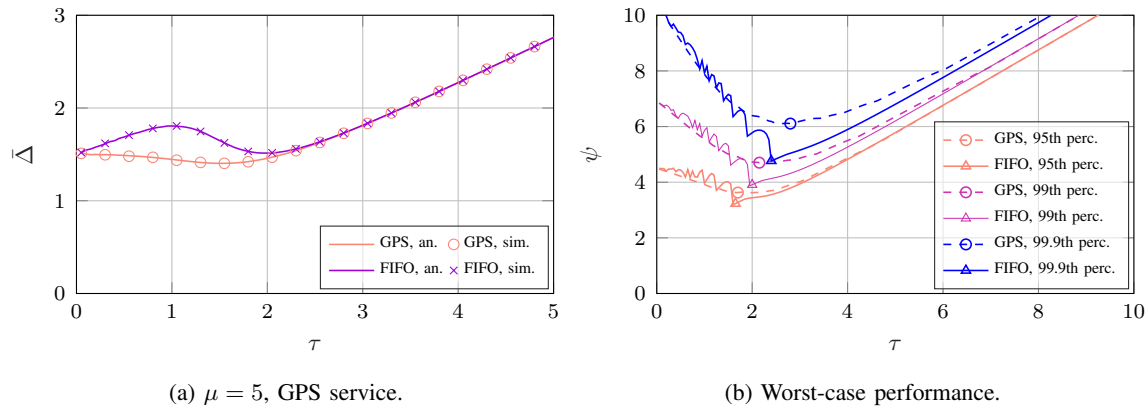

    \centering
\subfloat[$\mu=5$, \gls{gps} service.\label{fig:expage_tau}]
{\input{./fig/exp_aoi_tau.tex}}
\subfloat[Worst-case performance.\label{fig:perc_tau}]
{\input{./fig/aoi_perc_tau.tex}}
    \caption{Worst-case \gls{paoi} performance of the \gls{gps} and \gls{fifo} systems as a function of $\tau$, with $N=6$, $\mu=4$, and $B=6$ with identical inter-batch times. The mark indicates the optimal value of $\tau$ for each percentile and service policy.}
    \label{fig:aoi_tau}
\end{figure*}

We can then look at the expected and worst-case age for a fixed scenario, with $N=6$, $\mu=4$, and individual clients spaced out by a constant period $\nu$, as a function of the overall period $\tau=N\nu$. We compute the expected \gls{aoi} and the 95th, 99th, and 99.9th percentiles of the \gls{paoi} distribution as a function of the frame period $\tau$, as shown in Fig.~\ref{fig:aoi_tau}. We first analyze the expected \gls{aoi}, shown in Fig.~\ref{fig:expage_tau}: if the value of $\tau$ is high, i.e., if the load on the system is low, both systems perform equally as well, as we discussed when considering the \gls{cdf} plots. However, the expected age for the \gls{gps} system follows the classic U-shaped curve, while the \gls{fifo} system performs worse when the load is higher, with a peak around $\tau\simeq1$. This effect is due to the higher latency that successful frames in a \gls{fifo} system experience: while the success probability for a frame is generally similar for the two systems, frames served with \gls{gps} often have a shorter latency, reducing the \gls{aoi} significantly when most frames are successful (i.e., when the latency is a significant component of the \gls{aoi}). The effect becomes smaller as the frame period gets shorter, i.e., as latency becomes less relevant in determining the \gls{aoi}.

We can then consider the worst-case performance, shown in Fig.~\ref{fig:perc_tau}: in general, \gls{fifo} systems provide a better performance for intermediate loads, particularly when considering the extreme percentiles, while the performance is similar at both very high and very low loads. Interestingly, the percentiles of the age for the \gls{fifo} system show some oscillations for higher loads, instead of decreasing monotonically with $\tau$ as for the \gls{gps} system: this might be due to the discrete nature of arrivals to the system, which results in the staircase-shaped \gls{cdf}, causing small perturbations to the system to have stronger effects.

\begin{figure*}[t!]
    \centering
\subfloat[\gls{gps} service.\label{fig:gps_spread}]
{\begin{tikzpicture}
\begin{axis}[%
width=\tfwidth,
height=\sfheight,
xmin=0,
xmax=1,
legend style={legend cell align=left, fill opacity=0.6, draw opacity=1, text opacity=1, legend columns=2, align=left, draw=white!15!black, font=\tiny, at={(0.985, 0.985)}, anchor=north east},
xlabel style={font=\footnotesize\color{white!15!black}},
xlabel={$\xi$},
ymin=3,
ymax=5,
xmajorgrids,
ymajorgrids,
ylabel style={font=\footnotesize\color{white!15!black}},
ylabel={$\psi$},
axis background/.style={fill=white}
]
\addplot [color=color1]
  table[row sep=crcr]{%
0	3.697\\
0.01	3.697\\
0.02	3.696\\
0.03	3.695\\
0.04	3.694\\
0.05	3.693\\
0.06	3.692\\
0.07	3.692\\
0.08	3.691\\
0.09	3.69\\
0.1	3.689\\
0.11	3.688\\
0.12	3.687\\
0.13	3.686\\
0.14	3.686\\
0.15	3.685\\
0.16	3.684\\
0.17	3.683\\
0.18	3.682\\
0.19	3.681\\
0.2	3.68\\
0.21	3.679\\
0.22	3.678\\
0.23	3.678\\
0.24	3.677\\
0.25	3.676\\
0.26	3.675\\
0.27	3.674\\
0.28	3.673\\
0.29	3.672\\
0.3	3.671\\
0.31	3.67\\
0.32	3.669\\
0.33	3.669\\
0.34	3.667\\
0.35	3.667\\
0.36	3.666\\
0.37	3.666\\
0.38	3.665\\
0.39	3.664\\
0.4	3.664\\
0.41	3.663\\
0.42	3.662\\
0.43	3.662\\
0.44	3.661\\
0.45	3.66\\
0.46	3.66\\
0.47	3.659\\
0.48	3.658\\
0.49	3.658\\
0.5	3.657\\
0.51	3.656\\
0.52	3.656\\
0.53	3.655\\
0.54	3.654\\
0.55	3.654\\
0.56	3.653\\
0.57	3.652\\
0.58	3.652\\
0.59	3.651\\
0.6	3.65\\
0.61	3.65\\
0.62	3.649\\
0.63	3.648\\
0.64	3.648\\
0.65	3.647\\
0.66	3.646\\
0.67	3.646\\
0.68	3.645\\
0.69	3.644\\
0.7	3.643\\
0.71	3.643\\
0.72	3.642\\
0.73	3.641\\
0.74	3.641\\
0.75	3.64\\
0.76	3.639\\
0.77	3.639\\
0.78	3.638\\
0.79	3.637\\
0.8	3.637\\
0.81	3.636\\
0.82	3.635\\
0.83	3.635\\
0.84	3.634\\
0.85	3.633\\
0.86	3.633\\
0.87	3.632\\
0.88	3.631\\
0.89	3.631\\
0.9	3.63\\
0.91	3.629\\
0.92	3.629\\
0.93	3.628\\
0.94	3.627\\
0.95	3.627\\
0.96	3.626\\
0.97	3.625\\
0.98	3.625\\
0.99	3.624\\
1	3.623\\
};
\addlegendentry{Client 1}

\addplot [color=color2,semithick]
  table[row sep=crcr]{%
0	3.697\\
0.01	3.696\\
0.02	3.694\\
0.03	3.693\\
0.04	3.691\\
0.05	3.69\\
0.06	3.688\\
0.07	3.687\\
0.08	3.686\\
0.09	3.684\\
0.1	3.683\\
0.11	3.681\\
0.12	3.68\\
0.13	3.679\\
0.14	3.677\\
0.15	3.676\\
0.16	3.674\\
0.17	3.673\\
0.18	3.672\\
0.19	3.671\\
0.2	3.669\\
0.21	3.668\\
0.22	3.667\\
0.23	3.666\\
0.24	3.664\\
0.25	3.663\\
0.26	3.662\\
0.27	3.661\\
0.28	3.66\\
0.29	3.658\\
0.3	3.657\\
0.31	3.656\\
0.32	3.655\\
0.33	3.654\\
0.34	3.653\\
0.35	3.652\\
0.36	3.651\\
0.37	3.65\\
0.38	3.649\\
0.39	3.648\\
0.4	3.647\\
0.41	3.646\\
0.42	3.645\\
0.43	3.644\\
0.44	3.643\\
0.45	3.643\\
0.46	3.642\\
0.47	3.641\\
0.48	3.64\\
0.49	3.639\\
0.5	3.638\\
0.51	3.638\\
0.52	3.637\\
0.53	3.636\\
0.54	3.636\\
0.55	3.635\\
0.56	3.634\\
0.57	3.634\\
0.58	3.633\\
0.59	3.632\\
0.6	3.632\\
0.61	3.631\\
0.62	3.631\\
0.63	3.63\\
0.64	3.629\\
0.65	3.629\\
0.66	3.628\\
0.67	3.628\\
0.68	3.628\\
0.69	3.627\\
0.7	3.627\\
0.71	3.626\\
0.72	3.626\\
0.73	3.626\\
0.74	3.625\\
0.75	3.625\\
0.76	3.625\\
0.77	3.624\\
0.78	3.624\\
0.79	3.624\\
0.8	3.624\\
0.81	3.623\\
0.82	3.623\\
0.83	3.623\\
0.84	3.623\\
0.85	3.623\\
0.86	3.623\\
0.87	3.623\\
0.88	3.623\\
0.89	3.623\\
0.9	3.623\\
0.91	3.623\\
0.92	3.623\\
0.93	3.623\\
0.94	3.623\\
0.95	3.623\\
0.96	3.623\\
0.97	3.623\\
0.98	3.623\\
0.99	3.623\\
1	3.623\\
};
\addlegendentry{Client 2}

\addplot [color=color3,semithick]
  table[row sep=crcr]{%
0	3.643\\
0.01	3.642\\
0.02	3.641\\
0.03	3.641\\
0.04	3.64\\
0.05	3.64\\
0.06	3.639\\
0.07	3.639\\
0.08	3.638\\
0.09	3.638\\
0.1	3.638\\
0.11	3.637\\
0.12	3.637\\
0.13	3.636\\
0.14	3.636\\
0.15	3.635\\
0.16	3.635\\
0.17	3.634\\
0.18	3.634\\
0.19	3.634\\
0.2	3.633\\
0.21	3.633\\
0.22	3.632\\
0.23	3.632\\
0.24	3.631\\
0.25	3.631\\
0.26	3.63\\
0.27	3.63\\
0.28	3.629\\
0.29	3.629\\
0.3	3.629\\
0.31	3.628\\
0.32	3.628\\
0.33	3.627\\
0.34	3.627\\
0.35	3.626\\
0.36	3.626\\
0.37	3.625\\
0.38	3.625\\
0.39	3.625\\
0.4	3.624\\
0.41	3.624\\
0.42	3.623\\
0.43	3.623\\
0.44	3.622\\
0.45	3.622\\
0.46	3.622\\
0.47	3.621\\
0.48	3.621\\
0.49	3.62\\
0.5	3.62\\
0.51	3.619\\
0.52	3.619\\
0.53	3.619\\
0.54	3.618\\
0.55	3.618\\
0.56	3.617\\
0.57	3.617\\
0.58	3.617\\
0.59	3.616\\
0.6	3.616\\
0.61	3.615\\
0.62	3.615\\
0.63	3.615\\
0.64	3.614\\
0.65	3.614\\
0.66	3.614\\
0.67	3.613\\
0.68	3.613\\
0.69	3.612\\
0.7	3.612\\
0.71	3.612\\
0.72	3.611\\
0.73	3.611\\
0.74	3.611\\
0.75	3.61\\
0.76	3.61\\
0.77	3.61\\
0.78	3.609\\
0.79	3.609\\
0.8	3.609\\
0.81	3.609\\
0.82	3.608\\
0.83	3.609\\
0.84	3.609\\
0.85	3.61\\
0.86	3.611\\
0.87	3.612\\
0.88	3.613\\
0.89	3.613\\
0.9	3.614\\
0.91	3.615\\
0.92	3.616\\
0.93	3.617\\
0.94	3.618\\
0.95	3.619\\
0.96	3.619\\
0.97	3.62\\
0.98	3.621\\
0.99	3.622\\
1	3.623\\
};
\addlegendentry{Client 3}

\addplot [color=color4]
  table[row sep=crcr]{%
0	3.645\\
0.01	3.644\\
0.02	3.644\\
0.03	3.644\\
0.04	3.643\\
0.05	3.643\\
0.06	3.643\\
0.07	3.643\\
0.08	3.642\\
0.09	3.642\\
0.1	3.642\\
0.11	3.641\\
0.12	3.641\\
0.13	3.641\\
0.14	3.641\\
0.15	3.64\\
0.16	3.64\\
0.17	3.64\\
0.18	3.639\\
0.19	3.639\\
0.2	3.639\\
0.21	3.639\\
0.22	3.638\\
0.23	3.638\\
0.24	3.638\\
0.25	3.638\\
0.26	3.637\\
0.27	3.637\\
0.28	3.637\\
0.29	3.636\\
0.3	3.636\\
0.31	3.636\\
0.32	3.636\\
0.33	3.635\\
0.34	3.635\\
0.35	3.635\\
0.36	3.635\\
0.37	3.634\\
0.38	3.634\\
0.39	3.634\\
0.4	3.634\\
0.41	3.633\\
0.42	3.633\\
0.43	3.633\\
0.44	3.633\\
0.45	3.632\\
0.46	3.632\\
0.47	3.632\\
0.48	3.632\\
0.49	3.631\\
0.5	3.631\\
0.51	3.631\\
0.52	3.631\\
0.53	3.63\\
0.54	3.63\\
0.55	3.63\\
0.56	3.63\\
0.57	3.629\\
0.58	3.629\\
0.59	3.629\\
0.6	3.629\\
0.61	3.629\\
0.62	3.628\\
0.63	3.628\\
0.64	3.628\\
0.65	3.628\\
0.66	3.628\\
0.67	3.627\\
0.68	3.627\\
0.69	3.627\\
0.7	3.627\\
0.71	3.627\\
0.72	3.626\\
0.73	3.626\\
0.74	3.626\\
0.75	3.626\\
0.76	3.626\\
0.77	3.626\\
0.78	3.625\\
0.79	3.625\\
0.8	3.625\\
0.81	3.625\\
0.82	3.625\\
0.83	3.625\\
0.84	3.625\\
0.85	3.624\\
0.86	3.624\\
0.87	3.624\\
0.88	3.624\\
0.89	3.624\\
0.9	3.624\\
0.91	3.624\\
0.92	3.624\\
0.93	3.624\\
0.94	3.624\\
0.95	3.624\\
0.96	3.623\\
0.97	3.623\\
0.98	3.623\\
0.99	3.623\\
1	3.623\\
};
\addlegendentry{Client 4}

\addplot [color=color5,semithick]
  table[row sep=crcr]{%
0	3.651\\
0.01	3.65\\
0.02	3.649\\
0.03	3.649\\
0.04	3.649\\
0.05	3.649\\
0.06	3.648\\
0.07	3.648\\
0.08	3.648\\
0.09	3.647\\
0.1	3.647\\
0.11	3.647\\
0.12	3.646\\
0.13	3.646\\
0.14	3.646\\
0.15	3.646\\
0.16	3.645\\
0.17	3.645\\
0.18	3.645\\
0.19	3.644\\
0.2	3.644\\
0.21	3.644\\
0.22	3.643\\
0.23	3.643\\
0.24	3.643\\
0.25	3.642\\
0.26	3.642\\
0.27	3.642\\
0.28	3.642\\
0.29	3.641\\
0.3	3.641\\
0.31	3.641\\
0.32	3.64\\
0.33	3.64\\
0.34	3.64\\
0.35	3.639\\
0.36	3.639\\
0.37	3.639\\
0.38	3.639\\
0.39	3.638\\
0.4	3.638\\
0.41	3.638\\
0.42	3.637\\
0.43	3.637\\
0.44	3.637\\
0.45	3.636\\
0.46	3.636\\
0.47	3.636\\
0.48	3.636\\
0.49	3.635\\
0.5	3.635\\
0.51	3.635\\
0.52	3.634\\
0.53	3.634\\
0.54	3.634\\
0.55	3.634\\
0.56	3.633\\
0.57	3.633\\
0.58	3.633\\
0.59	3.632\\
0.6	3.632\\
0.61	3.632\\
0.62	3.632\\
0.63	3.631\\
0.64	3.631\\
0.65	3.631\\
0.66	3.631\\
0.67	3.63\\
0.68	3.63\\
0.69	3.63\\
0.7	3.63\\
0.71	3.629\\
0.72	3.629\\
0.73	3.629\\
0.74	3.629\\
0.75	3.628\\
0.76	3.628\\
0.77	3.628\\
0.78	3.628\\
0.79	3.627\\
0.8	3.627\\
0.81	3.627\\
0.82	3.627\\
0.83	3.626\\
0.84	3.626\\
0.85	3.626\\
0.86	3.626\\
0.87	3.626\\
0.88	3.625\\
0.89	3.625\\
0.9	3.625\\
0.91	3.625\\
0.92	3.625\\
0.93	3.624\\
0.94	3.624\\
0.95	3.624\\
0.96	3.624\\
0.97	3.624\\
0.98	3.624\\
0.99	3.623\\
1	3.623\\
};
\addlegendentry{Client 5}

\addplot [color=color6,semithick]
  table[row sep=crcr]{%
0	3.662\\
0.01	3.661\\
0.02	3.661\\
0.03	3.66\\
0.04	3.66\\
0.05	3.66\\
0.06	3.659\\
0.07	3.659\\
0.08	3.659\\
0.09	3.658\\
0.1	3.658\\
0.11	3.657\\
0.12	3.657\\
0.13	3.657\\
0.14	3.656\\
0.15	3.656\\
0.16	3.656\\
0.17	3.655\\
0.18	3.655\\
0.19	3.654\\
0.2	3.654\\
0.21	3.654\\
0.22	3.653\\
0.23	3.653\\
0.24	3.652\\
0.25	3.652\\
0.26	3.652\\
0.27	3.651\\
0.28	3.651\\
0.29	3.651\\
0.3	3.65\\
0.31	3.65\\
0.32	3.649\\
0.33	3.649\\
0.34	3.649\\
0.35	3.648\\
0.36	3.648\\
0.37	3.647\\
0.38	3.647\\
0.39	3.647\\
0.4	3.646\\
0.41	3.646\\
0.42	3.645\\
0.43	3.645\\
0.44	3.645\\
0.45	3.644\\
0.46	3.644\\
0.47	3.643\\
0.48	3.643\\
0.49	3.643\\
0.5	3.642\\
0.51	3.642\\
0.52	3.641\\
0.53	3.641\\
0.54	3.641\\
0.55	3.64\\
0.56	3.64\\
0.57	3.639\\
0.58	3.639\\
0.59	3.639\\
0.6	3.638\\
0.61	3.638\\
0.62	3.637\\
0.63	3.637\\
0.64	3.637\\
0.65	3.636\\
0.66	3.636\\
0.67	3.636\\
0.68	3.635\\
0.69	3.635\\
0.7	3.634\\
0.71	3.634\\
0.72	3.634\\
0.73	3.633\\
0.74	3.633\\
0.75	3.632\\
0.76	3.632\\
0.77	3.632\\
0.78	3.631\\
0.79	3.631\\
0.8	3.63\\
0.81	3.63\\
0.82	3.63\\
0.83	3.629\\
0.84	3.629\\
0.85	3.629\\
0.86	3.628\\
0.87	3.628\\
0.88	3.627\\
0.89	3.627\\
0.9	3.627\\
0.91	3.626\\
0.92	3.626\\
0.93	3.626\\
0.94	3.625\\
0.95	3.625\\
0.96	3.625\\
0.97	3.624\\
0.98	3.624\\
0.99	3.624\\
1	3.623\\
};
\addlegendentry{Client 6}

\end{axis}
\end{tikzpicture}%
}
\subfloat[\gls{fifo} queuing.\label{fig:fifo_spread}]
{\begin{tikzpicture}
\begin{axis}[%
width=\tfwidth,
height=\sfheight,
xmin=0,
xmax=1,
legend style={legend cell align=left, fill opacity=0.6, draw opacity=1, text opacity=1, legend columns=2, align=left, draw=white!15!black, font=\tiny, at={(0.985, 0.985)}, anchor=north east},
xlabel style={font=\footnotesize\color{white!15!black}},
xlabel={$\xi$},
ymin=3,
ymax=5,
xmajorgrids,
ymajorgrids,
ylabel style={font=\footnotesize\color{white!15!black}},
ylabel={$\psi$},
axis background/.style={fill=white}
]
\addplot [color=color1]
  table[row sep=crcr]{%
0	4.614\\
0.01	4.608\\
0.02	4.602\\
0.03	4.596\\
0.04	4.59\\
0.05	4.584\\
0.06	4.577\\
0.07	4.571\\
0.08	4.565\\
0.09	4.558\\
0.1	4.552\\
0.11	4.545\\
0.12	4.539\\
0.13	4.532\\
0.14	4.525\\
0.15	4.519\\
0.16	4.512\\
0.17	4.505\\
0.18	4.498\\
0.19	4.491\\
0.2	4.484\\
0.21	4.476\\
0.22	4.469\\
0.23	4.462\\
0.24	4.454\\
0.25	4.446\\
0.26	4.439\\
0.27	4.431\\
0.28	4.423\\
0.29	4.415\\
0.3	4.407\\
0.31	4.399\\
0.32	4.39\\
0.33	4.382\\
0.34	4.373\\
0.35	4.364\\
0.36	4.355\\
0.37	4.346\\
0.38	4.337\\
0.39	4.327\\
0.4	4.318\\
0.41	4.308\\
0.42	4.298\\
0.43	4.288\\
0.44	4.277\\
0.45	4.266\\
0.46	4.255\\
0.47	4.244\\
0.48	4.233\\
0.49	4.221\\
0.5	4.209\\
0.51	4.196\\
0.52	4.184\\
0.53	4.17\\
0.54	4.157\\
0.55	4.142\\
0.56	4.127\\
0.57	4.112\\
0.58	4.096\\
0.59	4.079\\
0.6	4.06\\
0.61	4.041\\
0.62	4.021\\
0.63	3.998\\
0.64	3.974\\
0.65	3.947\\
0.66	3.916\\
0.67	3.881\\
0.68	3.838\\
0.69	3.779\\
0.7	3.679\\
0.71	3.299\\
0.72	3.297\\
0.73	3.295\\
0.74	3.293\\
0.75	3.29\\
0.76	3.288\\
0.77	3.286\\
0.78	3.283\\
0.79	3.281\\
0.8	3.279\\
0.81	3.276\\
0.82	3.274\\
0.83	3.272\\
0.84	3.269\\
0.85	3.267\\
0.86	3.265\\
0.87	3.262\\
0.88	3.26\\
0.89	3.258\\
0.9	3.256\\
0.91	3.253\\
0.92	3.251\\
0.93	3.249\\
0.94	3.246\\
0.95	3.244\\
0.96	3.242\\
0.97	3.239\\
0.98	3.237\\
0.99	3.235\\
1	3.232\\
};
\addlegendentry{Client 1}

\addplot [color=color2]
  table[row sep=crcr]{%
0	3.106\\
0.01	3.107\\
0.02	3.109\\
0.03	3.111\\
0.04	3.112\\
0.05	3.114\\
0.06	3.115\\
0.07	3.117\\
0.08	3.118\\
0.09	3.12\\
0.1	3.121\\
0.11	3.123\\
0.12	3.125\\
0.13	3.126\\
0.14	3.128\\
0.15	3.129\\
0.16	3.13\\
0.17	3.132\\
0.18	3.133\\
0.19	3.135\\
0.2	3.136\\
0.21	3.138\\
0.22	3.139\\
0.23	3.141\\
0.24	3.142\\
0.25	3.143\\
0.26	3.145\\
0.27	3.146\\
0.28	3.148\\
0.29	3.149\\
0.3	3.15\\
0.31	3.152\\
0.32	3.153\\
0.33	3.154\\
0.34	3.156\\
0.35	3.157\\
0.36	3.158\\
0.37	3.16\\
0.38	3.161\\
0.39	3.162\\
0.4	3.164\\
0.41	3.165\\
0.42	3.166\\
0.43	3.167\\
0.44	3.169\\
0.45	3.17\\
0.46	3.171\\
0.47	3.172\\
0.48	3.174\\
0.49	3.175\\
0.5	3.176\\
0.51	3.177\\
0.52	3.179\\
0.53	3.18\\
0.54	3.181\\
0.55	3.182\\
0.56	3.184\\
0.57	3.185\\
0.58	3.186\\
0.59	3.187\\
0.6	3.188\\
0.61	3.189\\
0.62	3.191\\
0.63	3.192\\
0.64	3.193\\
0.65	3.194\\
0.66	3.195\\
0.67	3.196\\
0.68	3.198\\
0.69	3.199\\
0.7	3.2\\
0.71	3.201\\
0.72	3.202\\
0.73	3.203\\
0.74	3.204\\
0.75	3.205\\
0.76	3.207\\
0.77	3.208\\
0.78	3.209\\
0.79	3.21\\
0.8	3.211\\
0.81	3.212\\
0.82	3.213\\
0.83	3.214\\
0.84	3.215\\
0.85	3.216\\
0.86	3.217\\
0.87	3.219\\
0.88	3.22\\
0.89	3.221\\
0.9	3.222\\
0.91	3.223\\
0.92	3.224\\
0.93	3.225\\
0.94	3.226\\
0.95	3.227\\
0.96	3.228\\
0.97	3.229\\
0.98	3.23\\
0.99	3.231\\
1	3.232\\
};
\addlegendentry{Client 2}

\addplot [color=color3]
  table[row sep=crcr]{%
0	3.163\\
0.01	3.164\\
0.02	3.165\\
0.03	3.166\\
0.04	3.167\\
0.05	3.168\\
0.06	3.169\\
0.07	3.17\\
0.08	3.17\\
0.09	3.171\\
0.1	3.172\\
0.11	3.173\\
0.12	3.174\\
0.13	3.175\\
0.14	3.176\\
0.15	3.177\\
0.16	3.178\\
0.17	3.178\\
0.18	3.179\\
0.19	3.18\\
0.2	3.181\\
0.21	3.182\\
0.22	3.183\\
0.23	3.184\\
0.24	3.184\\
0.25	3.185\\
0.26	3.186\\
0.27	3.187\\
0.28	3.188\\
0.29	3.188\\
0.3	3.189\\
0.31	3.19\\
0.32	3.191\\
0.33	3.191\\
0.34	3.192\\
0.35	3.193\\
0.36	3.194\\
0.37	3.194\\
0.38	3.195\\
0.39	3.196\\
0.4	3.197\\
0.41	3.197\\
0.42	3.198\\
0.43	3.199\\
0.44	3.199\\
0.45	3.2\\
0.46	3.201\\
0.47	3.202\\
0.48	3.202\\
0.49	3.203\\
0.5	3.204\\
0.51	3.204\\
0.52	3.205\\
0.53	3.206\\
0.54	3.206\\
0.55	3.207\\
0.56	3.208\\
0.57	3.208\\
0.58	3.209\\
0.59	3.209\\
0.6	3.21\\
0.61	3.211\\
0.62	3.211\\
0.63	3.212\\
0.64	3.213\\
0.65	3.213\\
0.66	3.214\\
0.67	3.214\\
0.68	3.215\\
0.69	3.216\\
0.7	3.216\\
0.71	3.217\\
0.72	3.217\\
0.73	3.218\\
0.74	3.218\\
0.75	3.219\\
0.76	3.22\\
0.77	3.22\\
0.78	3.221\\
0.79	3.221\\
0.8	3.222\\
0.81	3.222\\
0.82	3.223\\
0.83	3.223\\
0.84	3.224\\
0.85	3.225\\
0.86	3.225\\
0.87	3.226\\
0.88	3.226\\
0.89	3.227\\
0.9	3.227\\
0.91	3.228\\
0.92	3.228\\
0.93	3.229\\
0.94	3.229\\
0.95	3.23\\
0.96	3.23\\
0.97	3.231\\
0.98	3.231\\
0.99	3.232\\
1	3.232\\
};
\addlegendentry{Client 3}

\addplot [color=color4]
  table[row sep=crcr]{%
0	3.185\\
0.01	3.186\\
0.02	3.186\\
0.03	3.187\\
0.04	3.188\\
0.05	3.188\\
0.06	3.189\\
0.07	3.19\\
0.08	3.19\\
0.09	3.191\\
0.1	3.192\\
0.11	3.192\\
0.12	3.193\\
0.13	3.193\\
0.14	3.194\\
0.15	3.195\\
0.16	3.195\\
0.17	3.196\\
0.18	3.196\\
0.19	3.197\\
0.2	3.198\\
0.21	3.198\\
0.22	3.199\\
0.23	3.199\\
0.24	3.2\\
0.25	3.2\\
0.26	3.201\\
0.27	3.201\\
0.28	3.202\\
0.29	3.203\\
0.3	3.203\\
0.31	3.204\\
0.32	3.204\\
0.33	3.205\\
0.34	3.205\\
0.35	3.206\\
0.36	3.206\\
0.37	3.207\\
0.38	3.207\\
0.39	3.208\\
0.4	3.208\\
0.41	3.209\\
0.42	3.209\\
0.43	3.21\\
0.44	3.21\\
0.45	3.211\\
0.46	3.211\\
0.47	3.212\\
0.48	3.212\\
0.49	3.212\\
0.5	3.213\\
0.51	3.213\\
0.52	3.214\\
0.53	3.214\\
0.54	3.215\\
0.55	3.215\\
0.56	3.216\\
0.57	3.216\\
0.58	3.216\\
0.59	3.217\\
0.6	3.217\\
0.61	3.218\\
0.62	3.218\\
0.63	3.218\\
0.64	3.219\\
0.65	3.219\\
0.66	3.22\\
0.67	3.22\\
0.68	3.221\\
0.69	3.221\\
0.7	3.221\\
0.71	3.222\\
0.72	3.222\\
0.73	3.222\\
0.74	3.223\\
0.75	3.223\\
0.76	3.224\\
0.77	3.224\\
0.78	3.224\\
0.79	3.225\\
0.8	3.225\\
0.81	3.225\\
0.82	3.226\\
0.83	3.226\\
0.84	3.227\\
0.85	3.227\\
0.86	3.227\\
0.87	3.228\\
0.88	3.228\\
0.89	3.228\\
0.9	3.229\\
0.91	3.229\\
0.92	3.229\\
0.93	3.23\\
0.94	3.23\\
0.95	3.23\\
0.96	3.231\\
0.97	3.231\\
0.98	3.232\\
0.99	3.232\\
1	3.232\\
};
\addlegendentry{Client 4}

\addplot [color=color5]
  table[row sep=crcr]{%
0	3.197\\
0.01	3.198\\
0.02	3.198\\
0.03	3.199\\
0.04	3.199\\
0.05	3.2\\
0.06	3.2\\
0.07	3.201\\
0.08	3.201\\
0.09	3.202\\
0.1	3.202\\
0.11	3.203\\
0.12	3.203\\
0.13	3.203\\
0.14	3.204\\
0.15	3.204\\
0.16	3.205\\
0.17	3.205\\
0.18	3.206\\
0.19	3.206\\
0.2	3.207\\
0.21	3.207\\
0.22	3.207\\
0.23	3.208\\
0.24	3.208\\
0.25	3.209\\
0.26	3.209\\
0.27	3.209\\
0.28	3.21\\
0.29	3.21\\
0.3	3.211\\
0.31	3.211\\
0.32	3.211\\
0.33	3.212\\
0.34	3.212\\
0.35	3.212\\
0.36	3.213\\
0.37	3.213\\
0.38	3.214\\
0.39	3.214\\
0.4	3.214\\
0.41	3.215\\
0.42	3.215\\
0.43	3.215\\
0.44	3.216\\
0.45	3.216\\
0.46	3.216\\
0.47	3.217\\
0.48	3.217\\
0.49	3.217\\
0.5	3.218\\
0.51	3.218\\
0.52	3.218\\
0.53	3.219\\
0.54	3.219\\
0.55	3.219\\
0.56	3.22\\
0.57	3.22\\
0.58	3.22\\
0.59	3.221\\
0.6	3.221\\
0.61	3.221\\
0.62	3.222\\
0.63	3.222\\
0.64	3.222\\
0.65	3.223\\
0.66	3.223\\
0.67	3.223\\
0.68	3.223\\
0.69	3.224\\
0.7	3.224\\
0.71	3.224\\
0.72	3.225\\
0.73	3.225\\
0.74	3.225\\
0.75	3.225\\
0.76	3.226\\
0.77	3.226\\
0.78	3.226\\
0.79	3.227\\
0.8	3.227\\
0.81	3.227\\
0.82	3.227\\
0.83	3.228\\
0.84	3.228\\
0.85	3.228\\
0.86	3.228\\
0.87	3.229\\
0.88	3.229\\
0.89	3.229\\
0.9	3.23\\
0.91	3.23\\
0.92	3.23\\
0.93	3.23\\
0.94	3.231\\
0.95	3.231\\
0.96	3.231\\
0.97	3.231\\
0.98	3.232\\
0.99	3.232\\
1	3.232\\
};
\addlegendentry{Client 5}

\addplot [color=color6]
  table[row sep=crcr]{%
0	3.206\\
0.01	3.206\\
0.02	3.207\\
0.03	3.207\\
0.04	3.207\\
0.05	3.208\\
0.06	3.208\\
0.07	3.209\\
0.08	3.209\\
0.09	3.209\\
0.1	3.21\\
0.11	3.21\\
0.12	3.21\\
0.13	3.211\\
0.14	3.211\\
0.15	3.211\\
0.16	3.212\\
0.17	3.212\\
0.18	3.212\\
0.19	3.213\\
0.2	3.213\\
0.21	3.213\\
0.22	3.213\\
0.23	3.214\\
0.24	3.214\\
0.25	3.214\\
0.26	3.215\\
0.27	3.215\\
0.28	3.215\\
0.29	3.216\\
0.3	3.216\\
0.31	3.216\\
0.32	3.216\\
0.33	3.217\\
0.34	3.217\\
0.35	3.217\\
0.36	3.218\\
0.37	3.218\\
0.38	3.218\\
0.39	3.218\\
0.4	3.219\\
0.41	3.219\\
0.42	3.219\\
0.43	3.219\\
0.44	3.22\\
0.45	3.22\\
0.46	3.22\\
0.47	3.22\\
0.48	3.221\\
0.49	3.221\\
0.5	3.221\\
0.51	3.221\\
0.52	3.222\\
0.53	3.222\\
0.54	3.222\\
0.55	3.222\\
0.56	3.223\\
0.57	3.223\\
0.58	3.223\\
0.59	3.223\\
0.6	3.224\\
0.61	3.224\\
0.62	3.224\\
0.63	3.224\\
0.64	3.224\\
0.65	3.225\\
0.66	3.225\\
0.67	3.225\\
0.68	3.225\\
0.69	3.226\\
0.7	3.226\\
0.71	3.226\\
0.72	3.226\\
0.73	3.226\\
0.74	3.227\\
0.75	3.227\\
0.76	3.227\\
0.77	3.227\\
0.78	3.228\\
0.79	3.228\\
0.8	3.228\\
0.81	3.228\\
0.82	3.228\\
0.83	3.229\\
0.84	3.229\\
0.85	3.229\\
0.86	3.229\\
0.87	3.229\\
0.88	3.23\\
0.89	3.23\\
0.9	3.23\\
0.91	3.23\\
0.92	3.231\\
0.93	3.231\\
0.94	3.231\\
0.95	3.231\\
0.96	3.231\\
0.97	3.232\\
0.98	3.232\\
0.99	3.232\\
1	3.232\\
};
\addlegendentry{Client 6}

\end{axis}
\end{tikzpicture}%
}
\subfloat[JFI.\label{fig:jain_spread}]
{\begin{tikzpicture}

\begin{semilogyaxis}[%
width=\tfwidth,
height=\sfheight,
legend style={legend cell align=left, fill opacity=0.6, draw opacity=1, text opacity=1, legend columns=1, align=left, draw=white!15!black, font=\tiny, at={(0.985, 0.985)}, anchor=north east},
xlabel style={font=\footnotesize\color{white!15!black}},
xmin=0,
xmax=1,
xlabel={$\xi$},
ymin=1e-6,
ymax=1,
max space between ticks=20,
ylabel style={font=\footnotesize\color{white!15!black}},
ylabel={$1-\mc{J}$},
xmajorgrids,
ymajorgrids,
]
\addplot [color=color3,semithick,dashed]
  table[row sep=crcr]{%
0  3.8853e-05\\
0.01	3.96543047891385e-05\\
0.02	3.8317295700474e-05\\
0.03	3.68936985845059e-05\\
0.04	3.57614751794744e-05\\
0.05	3.4310230005552e-05\\
0.06	3.31853334180066e-05\\
0.07	3.25398176390657e-05\\
0.08	3.21785735889479e-05\\
0.09	3.05695152746921e-05\\
0.1	2.92393690813908e-05\\
0.11	2.83837583241464e-05\\
0.12	2.74013605819912e-05\\
0.13	2.66800458512684e-05\\
0.14	2.5671716027742e-05\\
0.15	2.53985103989551e-05\\
0.16	2.40110456793952e-05\\
0.17	2.33794031497814e-05\\
0.18	2.26308334528857e-05\\
0.19	2.175693384876e-05\\
0.2	2.07658388934329e-05\\
0.21	1.97204471218404e-05\\
0.22	1.97312486869272e-05\\
0.23	1.93643511745556e-05\\
0.24	1.84916615317832e-05\\
0.25	1.77275207841054e-05\\
0.26	1.75838535647621e-05\\
0.27	1.66062095092734e-05\\
0.28	1.61913779478384e-05\\
0.29	1.55161839018314e-05\\
0.3	1.45996238575252e-05\\
0.31	1.4246478599822e-05\\
0.32	1.35245928475358e-05\\
0.33	1.40361556656421e-05\\
0.34	1.27041741299072e-05\\
0.35	1.30183304741793e-05\\
0.36	1.22844495863417e-05\\
0.37	1.27241690049962e-05\\
0.38	1.20149439363226e-05\\
0.39	1.14467492829951e-05\\
0.4	1.16697578226299e-05\\
0.41	1.12268270225924e-05\\
0.42	1.1024513047686e-05\\
0.43	1.09312165937903e-05\\
0.44	1.07770000715712e-05\\
0.45	1.04118710982926e-05\\
0.46	1.03352274649637e-05\\
0.47	1.00872557694531e-05\\
0.48	9.53513201329415e-06\\
0.49	1.01683717791401e-05\\
0.5	9.5097540576683e-06\\
0.51	9.46109466160028e-06\\
0.52	9.36916559368672e-06\\
0.53	9.01242570383598e-06\\
0.54	8.97204533900631e-06\\
0.55	8.83916021487963e-06\\
0.56	8.85081441948099e-06\\
0.57	8.37998697544862e-06\\
0.58	8.39127361551562e-06\\
0.59	8.41960567210531e-06\\
0.6	7.84024120059623e-06\\
0.61	8.30480910385401e-06\\
0.62	7.83331078479943e-06\\
0.63	7.45001636126563e-06\\
0.64	7.92328039889156e-06\\
0.65	7.36846972859695e-06\\
0.66	7.02420018416916e-06\\
0.67	7.5130915622168e-06\\
0.68	6.97471529909066e-06\\
0.69	7.07154248535957e-06\\
0.7	6.56658731978954e-06\\
0.71	6.60155035780985e-06\\
0.72	6.71099859494984e-06\\
0.73	6.22253756488167e-06\\
0.74	6.27588303059312e-06\\
0.75	6.19372152455e-06\\
0.76	5.87116741046501e-06\\
0.77	5.92450988268212e-06\\
0.78	5.91823714268802e-06\\
0.79	5.57274780998984e-06\\
0.8	5.44433831928259e-06\\
0.81	5.20013150007159e-06\\
0.82	5.34485711600752e-06\\
0.83	4.76387561898495e-06\\
0.84	4.50844187582788e-06\\
0.85	3.89343809248377e-06\\
0.86	3.43663215118806e-06\\
0.87	2.90369135880209e-06\\
0.88	2.29503762494421e-06\\
0.89	2.29503762483318e-06\\
0.9	1.8635290507385e-06\\
0.91	1.41311291612656e-06\\
0.92	1.20357404076454e-06\\
0.93	8.8002288389788e-07\\
0.94	5.77568320392352e-07\\
0.95	4.48473394087934e-07\\
0.96	3.72386069646957e-07\\
0.97	1.8832622439291e-07\\
0.98	1.18486506828219e-07\\
0.99	3.59724300347253e-08\\
1	3.33066907387547e-16\\
};
\addlegendentry{GPS}

\addplot [color=color6,semithick]
  table[row sep=crcr]{%
0 0.0243\\
0.01	0.024097123797477\\
0.02	0.0238773983114337\\
0.03	0.0236499844924185\\
0.04	0.0234419824593046\\
0.05	0.0232170046122797\\
0.06	0.0229809264890372\\
0.07	0.0227497668822693\\
0.08	0.0225621424572888\\
0.09	0.0223111187834503\\
0.1	0.0221008226561702\\
0.11	0.021860139507758\\
0.12	0.0216493836648501\\
0.13	0.0214209137274834\\
0.14	0.0211757111825018\\
0.15	0.0209776265155431\\
0.16	0.0207447658106487\\
0.17	0.0205187756785601\\
0.18	0.0202954108962733\\
0.19	0.0200562921575571\\
0.2	0.0198275777152955\\
0.21	0.019577794012302\\
0.22	0.0193587659490583\\
0.23	0.0191249484947321\\
0.24	0.0188885662528726\\
0.25	0.0186459126586732\\
0.26	0.0184157800454802\\
0.27	0.0181828366227186\\
0.28	0.0179283393676569\\
0.29	0.0176917551810273\\
0.3	0.0174559897092924\\
0.31	0.0172129399294015\\
0.32	0.0169605704867686\\
0.33	0.0167229615282465\\
0.34	0.0164655053104891\\
0.35	0.016211197572865\\
0.36	0.0159524104417272\\
0.37	0.0157006544601376\\
0.38	0.0154518178923659\\
0.39	0.0151795524936694\\
0.4	0.0149267903285413\\
0.41	0.0146591627369973\\
0.42	0.0143995270605881\\
0.43	0.0141356432311621\\
0.44	0.0138432137086442\\
0.45	0.0135600906429225\\
0.46	0.0132856305569246\\
0.47	0.0130017905376891\\
0.48	0.0127258956731137\\
0.49	0.0124361172580452\\
0.5	0.0121378637052595\\
0.51	0.011837615372963\\
0.52	0.0115393656798689\\
0.53	0.0112126704780549\\
0.54	0.0109173739901852\\
0.55	0.0105825490041918\\
0.56	0.0102307417175433\\
0.57	0.00991097902095905\\
0.58	0.00956997062352161\\
0.59	0.009209561532185\\
0.6	0.008815790021373\\
0.61	0.00842961436088363\\
0.62	0.00803185337107504\\
0.63	0.00759238819568653\\
0.64	0.0071418830341442\\
0.65	0.00665221362592172\\
0.66	0.00611107098789399\\
0.67	0.00553469432311038\\
0.68	0.00485229005977272\\
0.69	0.00399021809097999\\
0.7	0.00272609896489884\\
0.71	9.37339010808058e-05\\
0.72	8.83083946501273e-05\\
0.73	8.28189988109074e-05\\
0.74	7.74784695386499e-05\\
0.75	7.04676377990543e-05\\
0.76	6.42958833371177e-05\\
0.77	6.00079600366898e-05\\
0.78	5.3721622937597e-05\\
0.79	4.93952905995521e-05\\
0.8	4.53111105526149e-05\\
0.81	4.02992386560053e-05\\
0.82	3.64258104427195e-05\\
0.83	3.29626445165765e-05\\
0.84	2.82720619887433e-05\\
0.85	2.52088006668094e-05\\
0.86	2.25440521073317e-05\\
0.87	1.81682683275675e-05\\
0.88	1.58391082264453e-05\\
0.89	1.35737503229594e-05\\
0.9	1.14428053547e-05\\
0.91	8.85156282282029e-06\\
0.92	7.25621912878172e-06\\
0.93	5.66826252168529e-06\\
0.94	3.9378879409524e-06\\
0.95	2.85304412228626e-06\\
0.96	1.9596415622436e-06\\
0.97	9.70515437903607e-07\\
0.98	4.6792654373462e-07\\
0.99	1.48885765938189e-07\\
1	0\\
};
\addlegendentry{FIFO}

\end{semilogyaxis}
\end{tikzpicture}%
}
    \caption{95th percentile of the \gls{paoi} as a function of the coherence $\xi$ with $N=6$, $\mu=4$, and $B=6$. The value of $\tau$ is optimized for each system.}
    \label{fig:batch_spread}
\end{figure*}
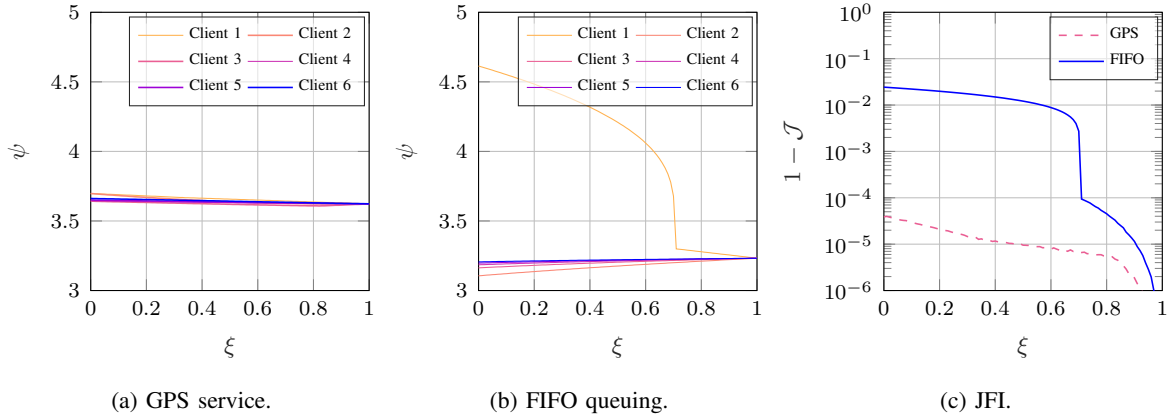

Finally, we consider the effect of \emph{drift} on the system: if one client loses synchronization and moves towards the next, while all others transmit at the allotted time, the effect is stronger on the \gls{fifo} system than on the \gls{gps} one. Considering the system with $N=6$, $B=6$ and equal values of $\nu$, i.e., clients arriving with a fixed $\frac{\tau}{N}$ interval, we suppose that client 1 starts to drift. We then define the coherence value $\xi\in[0,1]$, and consider a scenario in which $\nu_1=\xi\nu_3$ and $\nu_2=(2-\xi)\nu_3$. In other words, client 1 starts transmitting later and getting closer to client 2's allotted time. We then consider the 95th percentile of the \gls{paoi}, optimizing $\tau$ for the case with $\xi=1$ (i.e., not considering the drift), and compute the \gls{jfi} of the system, which is defined as:
\begin{equation}
 \mc{J}(x_1,\ldots,x_n)=\frac{\left(\sum_{i=1}^{n}x_i\right)^2}{n\sum_{i=1}^{n}x_i^2}.
\end{equation}
Fig.~\ref{fig:batch_spread} shows what happens in a \gls{gps} and \gls{fifo} system. In the former, whose performance is shown in Fig.~\ref{fig:gps_spread}, the 95th percentile \gls{paoi} increases slightly for all clients, and client 1 is noticeably underperforming relative to the others when $\xi$ is close to 0, but the change is minimal. On the other hand, the effect is much more noticeable in the \gls{fifo} system, shown in Fig.~\ref{fig:fifo_spread}: since client 1 spends a much shorter time at the head of the queue, its performance becomes much worse, and the \gls{paoi} increases by as much as 50\%. We can notice this from the \gls{jfi} plot, shown in Fig.~\ref{fig:jain_spread} on a logarithmic scale: while the fairness of the \gls{gps} system remains almost perfect, the \gls{fifo} system's fairness significantly degrades if $\xi<0.7$.

In conclusion, while the \gls{fifo} system can generally obtain better worst-case \gls{paoi} performance, it is also generally more sensitive to the parameters of the system, and slight timing inaccuracies can cause some of the clients to have significantly worse outcomes. The higher degree of randomness in the \gls{gps} scheme makes it more suitable for systems with a less precise synchronization, as it is generally more robust and can guarantee good performance even when the actual frame generation process is significantly altered.

\section{Conclusion}\label{sec:conc}

In this work, we studied the latency and \gls{paoi} performance of an edge computing server shared by $N$ clients, who generate periodic frames that must be processed in a timely fashion. We considered the well-known \gls{gps} and \gls{fifo} policies, deriving the expected \gls{aoi} and the latency and \gls{paoi} distributions for both. Our analysis yielded some interesting design insights, as the choice between the two service philosophies is non-trivial: in general, a \gls{fifo} system can provide better \gls{paoi} guarantees in the worst case if the frame generation period $\tau$ is controllable, but \gls{gps} works slightly better for higher levels of system load, can maintain a lower expected \gls{aoi}, and is generally more robust to perturbations in the arrival process and system parameters.

Future work on the subject may include a more realistic model in the same framework, considering different service time distributions and clients with different needs and frame processing complexity levels. Foresighted policies based on reinforcement learning, which can take the current state of the system into account and dynamically allocate resources to each client, are another interesting avenue of future research, which would bring the analytical results from this study closer to more complex network slicing scenarios, in which performance can only be measured empirically.

\bibliographystyle{IEEEtran}
\bibliography{bibliography.bib}

\end{document}